\documentclass[aps,prb,twocolumn,floatfix,superscriptaddress]{revtex4-1}
\usepackage{graphicx,txfonts,bm}
\usepackage[colorlinks,citecolor=magenta,linkcolor=blue]{hyperref}

\begin{document}


\title{Electronic structure of cerium: A comprehensive first-principles study}


\author{Li Huang}
\email{lihuang.dmft@gmail.com}
\affiliation{Science and Technology on Surface Physics and Chemistry Laboratory, P.O. Box 9-35, Jiangyou 621908, China}

\author{Haiyan Lu}
\affiliation{Science and Technology on Surface Physics and Chemistry Laboratory, P.O. Box 9-35, Jiangyou 621908, China}

\date{\today}


\begin{abstract}
Cerium, in which the 4$f$ valence electrons live at the brink between localized and itinerant characters, exhibits varying crystal structures and therefore anomalous physical properties with respect to temperature and pressure. Understanding its electronic structure and related lattice properties is one of the central topics in condensed matter theory. In the present work, we employed the state-of-the-art first-principles many-body approach (i.e., the density functional theory in combination with the single-site dynamical mean-field theory) to study its electronic structure thoroughly. The momentum-resolved spectral functions, total and $4f$ partial density of states, optical conductivities, self-energy functions, and atomic eigenstate histograms  for cerium's four allotropes under ambient pressure were calculated and analyzed carefully. The calculated results demonstrate that the 4$f$ electrons in the $\alpha$, $\beta$, $\gamma$, and $\delta$ phases are all correlated with heavily remormalized electron masses. In the $\alpha$ phase, the 4$f$ electrons tend to be itinerant, which cause strong hybridization between the 4$f$ and $spd$ bands and remarkable 4$f$ valence state fluctuation. While for the other phases, the 4$f$ electrons are close to be localized. Our calculated results support the Kondo volume collapse scenario for the cerium $\alpha-\gamma$ transition. Finally, we examined the site dependence of $4f$ electronic structure in the $\beta$ phase. The calculated results suggest that it doesn't exhibit a site selective 4$f$ localized state, contrary to previous prediction. 
\end{abstract}


\maketitle


\section{Introduction\label{sec:intro}}

\begin{figure}[t]
\centering
\includegraphics[width=\columnwidth]{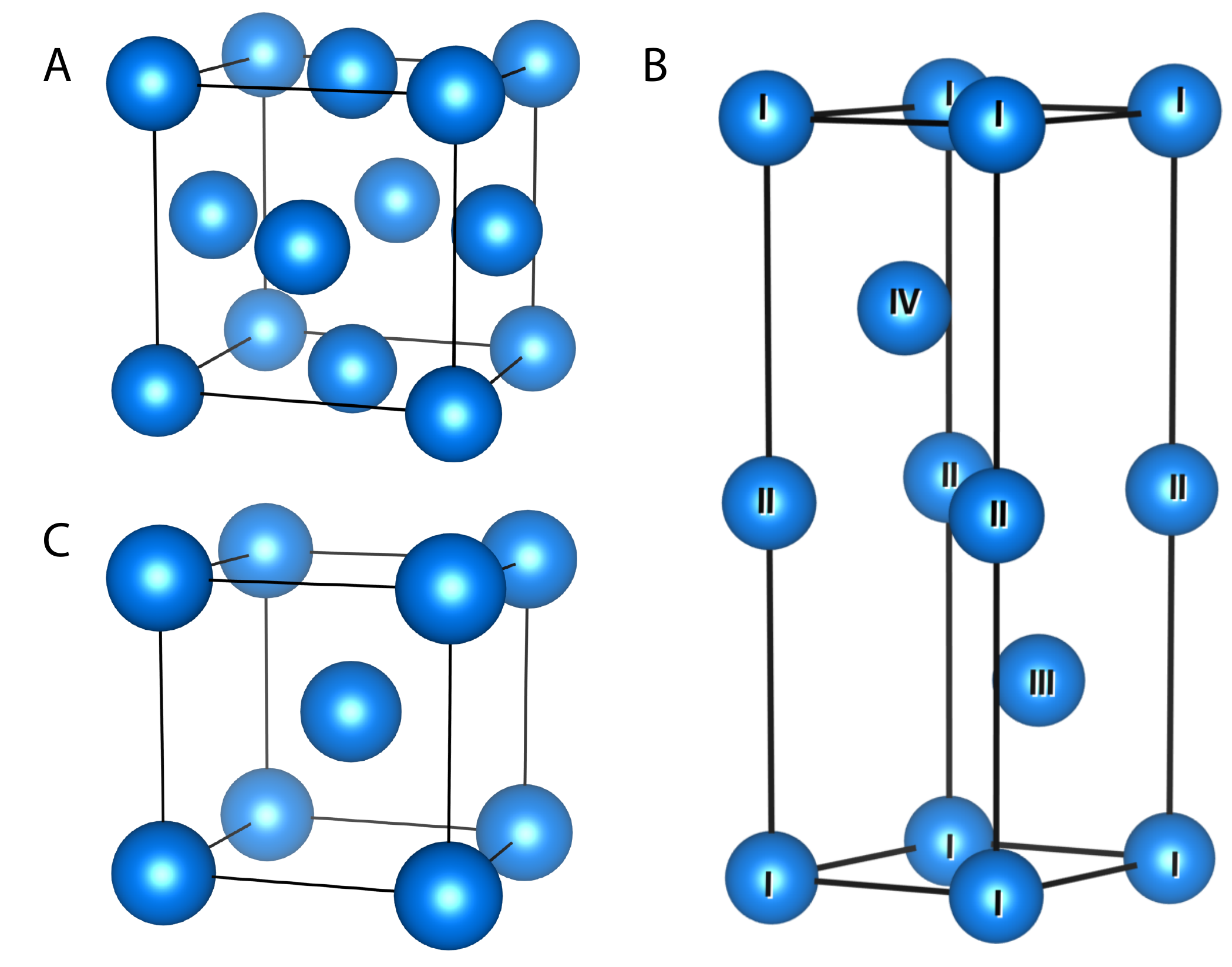}
\caption{(Color online). Crystal structures of cerium's allotropes studied in the present work~\cite{alex:2012}. (a) Face-centered cubic, $\alpha$ and $\gamma$ phases. (b) Double hexagonal close-packed, $\beta$ phase. (c) Body-centered cubic, $\delta$ phase. The different sites in the $\beta$ phase are depicted using Roman numerals (I $\sim$ IV). Actually, there are only two non-equivalent sites in $\beta$-Ce (Site 1: Ce$_{\text{I}}$ and Ce$_{\text{II}}$, Site 2: Ce$_{\text{III}}$ and Ce$_{\text{IV}}$)~\cite{koskenmaki1978337}. See the main text in Sec.~\ref{sec:discuss} for more details. \label{fig:tstruct}}
\end{figure}

Cerium (Ce) is an interesting rare earth element. It presents a rich mixture of lattice properties~\cite{koskenmaki1978337}. At ambient pressure, pure Ce metal may undergo three successive solid state phase transitions before reaching its liquid state, passing from $\alpha$, $\beta$, $\gamma$, to $\delta$ phase (see Fig.~\ref{fig:tstruct}). Under high pressure, three additional low-symmetry phases, $\alpha'$, $\alpha''$, and $\epsilon$, were identified~\cite{alex:2012}. One of unique physical properties of Ce is the huge volume expansion ($\sim$ 16\%) that occurs in the iso-structural $\alpha-\gamma$ phase transition~\cite{koskenmaki1978337}. The underlying mechanism and driving force of this transition are still in hot debate~\cite{bj:1974,PhysRevLett.74.2335,PhysRevLett.49.1106,PhysRevB.46.5047,PhysRevLett.92.105702,PhysRevLett.101.165703,PhysRevB.89.184426}. Upon heating, the $\beta-\gamma$ transformation involves a considerable and abnormal volume decrease ($\sim$ 1.2\%)~\cite{koskenmaki1978337}. The high-temperature phonon spectrum of $\gamma$-Ce shows pronounced phonon softening in the $L$ point in the $T[111]$ branch. The corresponding elastic properties are highly anisotropic too~\cite{krisch:2011,PhysRevB.19.5746,huang:2007,PhysRevB.25.6485}. 

The origins of these unusual properties of Ce have been intensively discussed before. In the first place, most of the investigations have attributed them to the Janus face of the $4f$ valence electrons~\cite{smith:198383}. When the $4f$ electrons are itinerant, they wander all over the lattice and take part in chemical bonding actively. In contrast, when they are localized, they tend to be inertial and form local magnetic moments. Coincidentally, the 4$f$ electrons in Ce pin in the intermediate regime of two limitations, i.e., fully localized and entirely itinerant. The dual nature of $4f$ valence electrons allows the physical properties of Ce to be easily tuned or modified via external conditions, such as temperature, pressure, chemical potential (impurity doping), and electromagnetic field etc. Secondly, it is no doubt that the $4f$ valence electrons in Ce are all correlated due to the strong Coulomb repulsive interactions among them. In addition, the spin-orbital coupling, crystal-field splitting, and lattice distortions effects in $f$-electron system also play essential roles. The interplay or entanglement between the Coulomb interaction, spin-orbital coupling, and lattice freedom of degrees affects the electronic structures in a subtle manner, and thus leads to unprecedentedly rich physics in Ce~\cite{koskenmaki1978337}.

In order to gain a deep insight into the intriguing properties of Ce, it is a high priority to understand its electronic structure at first. For one thing, Ce is a highly active metallic element. Thus, it is really a huge challenge to prepare suitable samples and carry out experimental researches. These obstacles have stimulated extensive theoretical studies on the electronic structure of Ce conversely. On the other side, Ce has been regarded as one of the hardest tentative systems for \emph{ab initio} electronic structures calculations, so numerous first-principles methods have been employed to make progress in the past decades. Nowadays, the equilibrium volumes $V_0$, bulk modulus $B$, elastic properties $C_{ij}$ of Ce have been well described by state-of-the-art density function theory (DFT) and its diverse extensions~\cite{PhysRev.140.A1133,PhysRev.136.B864}. Two of the most noteworthy achievements are as follows. Firstly, by using the density functional theory plus Gutzwiller variational method (dubbed as DFT + G), the equation of states and total energies of all major allotropes of Ce have been obtained~\cite{PhysRevB.90.161104,PhysRevLett.111.196801,PhysRevB.91.125148}. These data could be utilized to explore their phase transitions and phase instabilities, and finally depict the whole phase diagrams. Secondly, the density functional theory in combination with the single-site dynamical mean-field theory (dubbed as DFT + DMFT)~\cite{RevModPhys.78.865,RevModPhys.68.13,PhysRevB.81.195107} has been successfully applied to capture the evolution of 4$f$ electronic configurations and explain the enormous volume change between $\alpha$- and $\gamma$-Ce~\cite{Shorikov2015,PhysRevLett.87.276403,PhysRevLett.87.276404,PhysRevB.67.075108,PhysRevB.72.115125,PhysRevB.91.161103,PhysRevB.89.125110,PhysRevB.86.115116,PhysRevLett.115.256402}. Their thermodynamics~\cite{PhysRevB.89.195132,PhysRevLett.96.066402}, optical properties~\cite{PhysRevLett.94.036401,PhysRevB.81.195107}, and magnetic susceptibilities~\cite{PhysRevB.89.125113,PhysRevB.85.195109} were also well reproduced. 

However, despite great achievements have been obtained, many pivotal electronic properties of Ce remain untouched or unclear so far. To the authors' knowledge, besides $\alpha$- and $\gamma$-Ce, the electronic structures of the other phases of Ce have been rarely reported. These phases can be roughly classified into two categories: (1) Low-temperature $\beta$ phase, whose crystal structure includes multiple non-equivalent lattice sites. (2) High-temperature $\delta$ phase, which only stabilizes in a very narrow temperature range~\cite{high_pressure}. Actually, we know a little about their electronic structures and corresponding physical properties, including the optical and magnetic characters, lattice dynamics, and mechanical properties, etc. Besides, there are still some limitations and disadvantages in the first-principles approaches employed in the previous calculations. For examples, though the traditional DFT-based methods (such as the DFT + $U$ method) are capable of reproducing the lattice constants and bulk modulus of Ce, the price is to introduce some kind of artificial spin and/or orbital polarizations. Sometimes these assumptions are unphysical and in contrast with the experiments. The quasi-particle approximation (alias GWA) has been used to study the Ce $\alpha - \gamma$ phase transition~\cite{PhysRevLett.109.146402}. It yields better results, but there are some technical issues that need to be solved. Most of all, the GWA method is in essence based on a weak-coupling theory. Whether it can be used to study the strongly correlated systems is nonetheless questionable. The DFT + G method is able to describe simultaneously the ground state energetics and $f$ electronic structures of the major phases of Ce~\cite{PhysRevB.90.161104,PhysRevLett.111.196801,PhysRevB.91.125148}. Its theoretical framework is formalized on zero temperature, in other words, the temperature effect and thermal excitation are completely ignored. When this method is adopted to study the high-temperature phases of Ce, it is difficult to make an estimation about how large the deviation is. Similar to the GWA method, another drawback of the DFT + G method is that it is only suitable for the weak and intermediate correlated systems. Finally, the DFT + DMFT method can be used to capture the $4f$ electronic structure of Ce accurately under finite temperature and irrespective of the electronic correlation strength. But it is extremely time-consuming and technically complex so that it is almost an impossible mission to use it to scan the whole $P-T$ phase diagram of Ce. Virtually, only the $\alpha$ and $\gamma$ phases have been studied by using the DFT + DMFT method~\cite{PhysRevB.89.125113,PhysRevLett.94.036401,PhysRevLett.87.276403,PhysRevLett.87.276404,PhysRevB.67.075108,PhysRevB.72.115125,PhysRevB.89.195132,PhysRevLett.96.066402,PhysRevLett.115.256402,savrasov:2001,shim:2007,zhu:2013,PhysRevB.91.161103,PhysRevB.89.125110,PhysRevB.86.115116,PhysRevB.85.195109}.

In general, a convincing description of the $4f$ electronic structure of Ce based on first-principles calculations is still lacking. None of the previous works can provide an unified picture for the evolution of electronic structures of all phases of Ce with respect to temperature and crystal symmetry. Thus, in order to fill in this gap, the main purpose of the present work is to figure out the entire electronic structures of Ce by using the charge fully self-consistent DFT + DMFT method~\cite{RevModPhys.78.865}. Besides the high-pressure $\alpha'$-, $\alpha''$-, and $\epsilon$-Ce phases, the major allotropes of Ce were taken into considerations in the current calculations~\cite{high_pressure}. The momentum-resolved spectral functions, total and $4f$ partial density of states, valence state histograms, and optical properties were investigated thoroughly and compared with the experimental results if available. Our calculated results will shed new light on the fine electronic structures of Ce, and more importantly, enrich our understanding about the underlying $4f$ orbital physics in the lanthanides.

The rest of this paper is organized as follows. In Sec.~\ref{sec:method}, we firstly introduce the spirit and advantages of the DFT + DMFT computational framework. And then we elaborate the details for the DFT, DMFT, and post-processing calculations, respectively. Section~\ref{sec:results} is the major part of this paper. In this section we present the theoretical electronic structures and optical properties for the four allotropes of Ce under ambient pressure. Then the calculated results are compared with the available experimental and theoretical data. In Sec.~\ref{sec:discuss}, we analyze the influence of inequivalent Ce atoms on the $4f$ electronic structure of $\beta$-Ce. Finally, Section~\ref{sec:summary} serves as a brief conclusion and outlook.

\section{Methods\label{sec:method}}

\begin{table*}[ht]
\caption{Key parameters used in the present DFT + DMFT calculations. In this table, the settings for $k$-points ($k$-mesh), radius of Muffin-tin sphere ($R_{\text{MT}}$), size of basis set ($R_{\text{MT}}K_{\text{MAX}}$), exchange-correlation functional (XC), double-counting term (DC), Coulomb repulsive interaction ($U$), Hund's exchange interaction ($J_{\text{H}}$), spin-orbital coupling constant ($\lambda_{\text{SO}}$), system temperature ($T$), and number of Monte Carlo sweeps ($N_{\text{sweeps}}$) per DMFT iteration (one-shot CT-HYB quantum impurity solver calculation) are shown. Here PBE means the Perdew-Burke-Ernzerhof functional~\cite{PhysRevLett.77.3865} and FLL means the fully localized limit scheme~\cite{jpcm:1997}. See main text for more explanations. \label{tab:param}}
\begin{ruledtabular}
\begin{tabular}{ccccccccccc}
cases & 
$k$-mesh &
$R_{\text{MT}}$ &
$R_{\text{MT}}K_{\text{MAX}}$ &
XC &
DC &
$U$ &
$J_{\text{H}}$ &
$\lambda_{\text{SO}}$ &
$T$ &
$N_{\text{sweeps}}$ \\
\hline
$\alpha$-Ce   & $17 \times 17 \times 17$ & 2.50 & 7.0 & PBE & FLL & 6.0 eV & 0.7 eV & 0.0953 & 116\ K & 4.0 $\times 10^9$ \\
$\beta$-Ce    & $26 \times 26 \times 07$ & 2.50 & 7.0 & PBE & FLL & 6.0 eV & 0.7 eV & 0.0953 & 232\ K & 2.0 $\times 10^9$ \\
$\gamma$-Ce   & $17 \times 17 \times 17$ & 2.50 & 7.0 & PBE & FLL & 6.0 eV & 0.7 eV & 0.0953 & 580\ K & 4.0 $\times 10^9$ \\
$\delta$-Ce   & $17 \times 17 \times 17$ & 2.50 & 7.0 & PBE & FLL & 6.0 eV & 0.7 eV & 0.0953 & 1009\ K& 4.0 $\times 10^9$ \\
\end{tabular}
\end{ruledtabular}
\end{table*}

As mentioned above, the 4$f$ electrons in Ce are strongly correlated~\cite{alex:2012}. In order to reach a proper description for its electronic structure, the strong Coulomb interaction among the $4f$ electrons and the spin-orbital coupling effects have to be considered on the same footing. Obviously, the traditional DFT methods which based on single-particle approximation or independent-electron picture break down~\cite{PhysRev.140.A1133,PhysRev.136.B864}. The single-site dynamical mean-field theory (DMFT) is a powerful and non-perturbative many-body approach to treat the local interactions between electrons~\cite{RevModPhys.68.13}. It firstly assumes that the lattice self-energy is momentum-independent in a localized basis, i.e., $\Sigma(\mathbf{k},\omega) \cong \Sigma(\omega)$, which becomes exact when the dimension of lattice is infinite ($d = \infty$). Then the lattice model which is in general intractable is mapped to a quantum impurity model in a self-consistent manner. The resulting quantum impurity model should be solved analytically or numerically using various quantum impurity solvers. Naturally, one can combine the DFT method with the DMFT approach to deal with the strongly correlated problems in realistic materials. The general spirit of the DFT + DMFT approach is very simple: The Hamiltonian produced by the DFT calculation is supplemented with a Hubbard-like local interaction term which encapsulates the electronic corrections, then the obtained Hamiltonian is self-consistently solved within the framework of the DMFT approach~\cite{RevModPhys.78.865}. As a matter of fact, it is probably the most powerful established method to study the electronic structures of strongly correlated materials, and has been successfully applied in the investigations of many actinide and lanthanide systems (for some pioneering and impressive works, see Refs.~[\onlinecite{dai:2003}] and~[\onlinecite{savrasov:2001}]). Here we thus adopted this approach to investigate the electronic structure of Ce. In the following, we will put flesh on the bones of the computational details (see also Table~\ref{tab:param}).

\subsection{DFT calculations}

The major objective of the DFT calculation is to generate the Kohn-Sham single-particle Hamiltonian $\hat{H}_{\text{KS}}$. Here we used the \texttt{WIEN2k}~\cite{wien2k} code, which implements a full-potential linear augmented plane wave formalism (FLAPW), to carry out this task. In order to evaluate the exchange-correlation potential, we chose the generalized gradient approximation, specially, the Perdew-Burke-Ernzerhof functional~\cite{PhysRevLett.77.3865}. The spin-orbital coupling was explicitly included in the calculations in a second-variational procedure. The convergence criteria for charge and energy were $10^{-4}$ e and $10^{-4}$ Ry, respectively. The experimental crystal structures of Ce were used throughout the calculations~\cite{koskenmaki1978337,alex:2012}. The important computational parameters are summarized in Table~\ref{tab:param}.

\subsection{DMFT calculations}

In the DFT + DMFT approach, the Hamiltonian reads:
\begin{equation}
\label{eq:ham}
\hat{H}_{\text{DFT+DMFT}} = \hat{H}_{\text{KS}} + \hat{H}_{\text{INT}} + \hat{H}_{\text{SOI}} - \hat{\Sigma}_{\text{DC}}.
\end{equation}
Here $\hat{H}_{\text{INT}}$ is the Coulomb interaction term for the 4$f$ orbitals. It is parameterized with the Slater integrals $F^0$, $F^2$, $F^4$ and $F^6$. For the $f$ electronic systems, the following relations apply~\cite{PhysRevB.59.9903}:
\begin{equation}
U=F^0,\ J_{\text{H}}=\frac{2}{45}F^2+\frac{1}{33}F^4+\frac{50}{1287}F^6,
\end{equation}
and
\begin{equation}
F^4=\frac{451}{675}F^2,\ F^6=\frac{1001}{2025}F^2,
\end{equation}
where $U$ is the Coulomb interaction strength and $J_{\text{H}}$ is the Hund's exchange parameter. So, with $U$ and $J_{\text{H}}$ as input parameters, the Slater integrals and hence $\hat{H}_{\text{INT}}$ are easily determined. The detailed values of $U$ and $J_{\text{H}}$ for Ce are summarized in Table~\ref{tab:param}, which are slightly larger than the values used in the previous calculations~\cite{PhysRevLett.115.256402}. The $\hat{H}_{\text{SOI}}$ term represents the spin-orbital interaction:
\begin{equation}
\hat{H}_{\text{SOI}} = \lambda_{\text{SO}} \sum_i \mathbf{l}_i \cdot \mathbf{s}_i,
\end{equation}
where $\lambda_{\text{SO}}$ is the strength for spin-orbital interaction. Its value is deduced from an individual DFT run. $\hat{\Sigma}_{\text{DC}}$ is the double-counting term for self-energy function. We used the fully localized limit scheme~\cite{jpcm:1997} in the present calculations:
\begin{equation}
\label{eq:dc}
\hat{\Sigma}_{\text{DC}} = U\Big(N_{f}-\frac{1}{2}\Big) - \frac{J_{\text{H}}}{2} \Big(N_{f}-1\Big),
\end{equation}
where the total 4$f$ occupancy $N_f$ was updated dynamically during the DFT + DMFT iterations. In the first DFT + DMFT iteration, $N_f$ was set to be 1.0. 

We employed the \texttt{EDMFTF} code~\cite{PhysRevB.81.195107}, which implements the DFT + DMFT computational engine and the corresponding quantum impurity solvers, to study the obtained DFT + DMFT Hamiltonian [see Eq.~(\ref{eq:ham})]. The temperature parameters $\beta$ ($\equiv 1/T$) were carefully chosen so that they are exactly in the temperature range where the allotropes of Ce usually live (see Table~\ref{tab:param}). The constructed multi-orbital quantum impurity models were solved using the hybridization expansion continuous-time quantum Monte Carlo impurity solver (dubbed as CT-HYB)~\cite{RevModPhys.83.349,PhysRevLett.97.076405,PhysRevB.75.155113}. Solving such seven-band impurity models with both general Coulomb interactions and spin-orbital coupling are incredibly difficult. The following strategies were exploited to simplify the calculations. First of all, there are multiple non-equivalent Ce atoms in $\beta$-Ce~\cite{koskenmaki1978337}. In principle, we have to treat them individually~\cite{zhu:2013}. In order to reduce the impurity problems we have to solve, these atoms were assumed to be equivalent. In other words, we ignored the site dependence of impurity problems. Latter we will make a detailed discussion about this aggressive simplification. Secondly, in the present work we not only utilized the good quantum numbers $N$ ($4f$ occupancy) and $J_z$ ($z$-component of total angular momentum) to reduce the sizes of matrix blocks of the local Hamiltonian, but also made a truncation for the local Hilbert space~\cite{PhysRevB.75.155113}. We only kept the atomic eigenstates with $N \in$ [0,3]. Finally, the lazy trace evaluation trick~\cite{PhysRevB.90.075149} was applied to accelerate the Monte Carlo sampling procedure further.

We performed charge fully self-consistent DFT + DMFT calculations, i.e., the electronic-correlation-corrected density matrix $\rho$ was built in the DMFT part, and then fed back to the DFT part to generate a new Kohn-Sham Hamiltonian $\hat{H}_{\text{KS}}$. Of the order of 40 DFT + DMFT iterations (which include about 800 DFT iterations and 40 one-shot DMFT calculations) were required to reach good convergence for the chemical potential $\mu$, charge density $\rho$, and total energy $E_{\text{DFT + DMFT}}$. The Matsubara self-energy functions $\Sigma(i\omega_n)$ generated in the last 10 DFT + DMFT iterations were collected and averaged for further post-processing.

\subsection{Analytical continuations and physical quantities}

The self-energy function on real axis $\Sigma(\omega)$ is an essential input for the calculations of some observable quantities, such as the momentum-resolved spectral function $A(\mathbf{k},\omega)$ and optical conductivity $\sigma(\omega)$. Unfortunately, the CT-HYB impurity solver generally works on imaginary-time axis~\cite{PhysRevLett.97.076405,PhysRevB.75.155113,RevModPhys.83.349}. Its direct output is the Matsubara self-energy function $\Sigma(i\omega_n)$ and impurity Green's function $G(i\omega_n)$, where $\omega_n$ is the Matsubara frequency with $\omega_n \equiv (2n+1) \pi / \beta$. So that we have to convert $\Sigma(i\omega_n)$ to $\Sigma(\omega)$ via elaborately analytical continuation operation at first. Since the original data for $\Sigma(i\omega_n)$ are always full of numerical fluctuation and noise, we adopt the following procedure~\cite{PhysRevB.81.195107} to tackle this problem. First, we define the auxiliary Green's function $\tilde{G}(i\omega_n)$ as follows:
\begin{equation}
\label{eq:tilde_g}
\tilde{G}(i\omega_n) = \frac{1}{i\omega_n - \Sigma(i\omega_n) + \Sigma(\infty)},
\end{equation}
where $\Sigma(\infty)$ means the high-frequency tail of the self-energy. Second, we perform analytical continuation for $\tilde{G}(i\omega_n)$ using the well-known maximum entropy method~\cite{jarrell}. The output is an auxiliary spectral function $\Im\tilde{G}(\omega)$. Third, we resort to the Kramers-Kronig relations to evaluate $\Re \tilde{G}(\omega)$ from $\Im\tilde{G}(\omega)$:
\begin{equation}
\Re \tilde{G}(\omega) = -\frac{1}{\pi} \int \frac{\Im \tilde{G}(\omega')}{\omega - \omega'}\text{d}\omega'.
\end{equation}
Finally, the desired $\Sigma(\omega)$ can be calculated directly using the reverse equation of Eq.~(\ref{eq:tilde_g}):
$\Sigma(\omega) = \omega - \tilde{G}^{-1}(\omega) + \Sigma(\infty)$.

Once $\Sigma(\omega)$ is ready, we then make use of it to calculate the momentum-resolved spectral functions $A(\mathbf{k},\omega)$ and integrated spectral functions $A(\omega)$ via the following equations:
\begin{equation}
\label{eq:akw}
A(\mathbf{k},\omega) = -\frac{1}{\pi}\Im\frac{1}{(\omega + \mu) \hat{\mathbf{I}} - \hat{H}_{\text{KS}}(\mathbf{k}) - \hat{E}(\mathbf{k})[ \Sigma(\omega) - \Sigma_{\text{DC}}]},
\end{equation}
\begin{equation}
A(\omega) = \int_{\Omega}  A(\mathbf{k},\omega) \text{d} \mathbf{k}.
\end{equation}
Here $\hat{\mathbf{I}}$ is the identity matrix, $\hat{H}_{\text{KS}}(\mathbf{k})$ the Kohn-Sham Hamiltonian, $\hat{E}(\mathbf{k})$ the embedding projector in momentum space~\cite{PhysRevB.81.195107}.

It is not a trivial task to calculate the optical conductivity $\sigma_{ab}(\omega)$. At first, we have to solve the following non-Hermitian eigenvalue problems for all momentum and frequency points $(\mathbf{k},\omega)$ to obtain both the eigenvalues and eigenvectors~\cite{PhysRevB.81.195107}:
\begin{equation}
\label{eq:eigen}
\big\{\hat{H}_{\text{KS}}(\mathbf{k}) + \hat{E}(\mathbf{k})[\Sigma(\omega) - \Sigma_{\text{DC}}]\big\} \psi_{\mathbf{k}\omega} = \epsilon_{\mathbf{k}\omega} \psi_{\mathbf{k}\omega}. 
\end{equation}
Then the optical conductivity is expressed as follows~\cite{PhysRevLett.94.036401}:
\begin{widetext}
\begin{equation}
\label{eq:optic}
\sigma_{ab}(\omega) = \frac{\pi e^2}{\omega} \sum_{ss' = \pm 1} ss' \sum_{\mathbf{k}}\sum_{jj'} \int^{+\omega/2}_{-\omega/2}
\text{d}\varepsilon
\frac{f(\varepsilon^{+}) - f(\varepsilon^{-})}{\omega}
\frac{\mathcal{M}^{ss',ab}_{\mathbf{k}jj'}(\varepsilon^{-},\varepsilon^{+})}{\omega + \epsilon^s_{\mathbf{k}j\varepsilon^{-}} - \epsilon^{s'}_{\mathbf{k}j'\varepsilon^{+}}}
\left(\frac{1}{\varepsilon^{-} + \mu - \epsilon^{s}_{\mathbf{k}j\varepsilon^{-}}} - \frac{1}{\varepsilon^{+} + \mu - \epsilon^{s'}_{\mathbf{k}j'\varepsilon^{+}}}\right),
\end{equation} 
\end{widetext}
where $a$, $b$, $j$, and $j'$ are orbital indices, and $f(\varepsilon)$ means the Fermi-Dirac distribution function. In Eq.~(\ref{eq:optic}), we have denoted $\varepsilon^{\pm} = \varepsilon \pm \omega/2$, and used the abbreviated notations $\epsilon^{+}_{\mathbf{k}j\varepsilon} = \epsilon_{\mathbf{k}j\varepsilon}$, $\epsilon^{-}_{\mathbf{k}j\varepsilon} = \epsilon^{*}_{\mathbf{k}j\varepsilon}$. The transition matrix elements $\mathcal{M}_{\mathbf{k}jj'}$ can be evaluated using the eigenvectors $\psi_{\mathbf{k}\omega}$ only [see Eq.~(\ref{eq:eigen})]. They are actually generalizations of the standard dipole allowed transition probabilities~\cite{PhysRevB.70.125112}. Note that in the present calculations the high-order vertex corrections to the conductivity are completely ignored~\cite{PhysRevLett.94.036401}.


\section{Results\label{sec:results}}

\begin{table}
\caption{The equilibrium lattice constants $a_0$ and bulk modulus $B$ for $\gamma$-Ce. The units for $a_0$ and $B$ are \AA\ and GPa, respectively. \label{tab:bulk}}
\begin{ruledtabular}
\begin{tabular}{rcc}
method & $a_0$ & $B$ \\
\hline
DFT + DMFT\footnotemark[1]  & 5.21 & 21.3 \\
DFT + DMFT\footnotemark[2]  & 5.07 & 36.0 \\
DFT + DMFT\footnotemark[3]  & 5.23 & 21.2 \\
DFT + $U$\footnotemark[4]   & 5.07 & 32.0 \\
DFT + $U$\footnotemark[5]   & 5.13 & 25.4 \\
DFT + $U$\footnotemark[6]   & 5.04 & 34.0 \\
PBE0\footnotemark[7]     & 5.22 & 28.3 \\
GGA-PBE ($4f$ in valence state)\footnotemark[7] & 4.68 & 36.6 \\
GGA-PBE ($4f$ in core state)\footnotemark[7]    & 5.30 & 28.8 \\
SIC-LSDA\footnotemark[8]     & 5.14 & 34.0 \\ 
PP-PW (4$f$ in core state)\footnotemark[9]      & 5.26 & 30.2 \\
VMC ($T = 0$\ K)\footnotemark[10]               & 4.99 & 38.0 \\
HP-XRD\footnotemark[11] & 5.16 & 21.0 \\
INS\footnotemark[12] &      & 14.8 \\
INS and XRD\footnotemark[13] & 5.17 & 19.0 \\ 
\end{tabular}
\end{ruledtabular}
\footnotetext[1]{The present work.}
\footnotetext[2]{With Hubbard-I impurity solver. See Ref.~[\onlinecite{amadon:2012}].}
\footnotetext[3]{With Hirsch-Fye quantum Monte Carlo impurity solver. See Ref.~[\onlinecite{PhysRevB.72.115125}].}
\footnotetext[4]{See Ref.~[\onlinecite{amadon:2012}].}
\footnotetext[5]{See Ref.~[\onlinecite{Hu2011669}].}
\footnotetext[6]{See Ref.~[\onlinecite{PhysRevB.77.155104}].}
\footnotetext[7]{Hybrid functional method. See Ref.~[\onlinecite{PhysRevB.93.075153}]}
\footnotetext[8]{Self-interaction correction plus local spin density approximation. See Ref.~[\onlinecite{PhysRevB.53.4275}]}
\footnotetext[9]{Pseudopotential plane-wave method. See Ref.~[\onlinecite{huang:2007}].}
\footnotetext[10]{Variational Monte Carlo method. See Ref.~[\onlinecite{PhysRevB.91.081101}].}
\footnotetext[11]{High-pressure X-ray diffraction. See Ref.~[\onlinecite{olsen:1985}].}
\footnotetext[12]{Inelastic neutron scattering. See Refs.~[\onlinecite{PhysRevB.19.5746}] and [\onlinecite{PhysRevB.25.6485}].}
\footnotetext[13]{Inelastic neutron scattering and X-ray power diffraction. See Ref.~[\onlinecite{PhysRevLett.92.105702}].}
\end{table}

\begin{figure*}[ht]
\centering
\includegraphics[width=\textwidth]{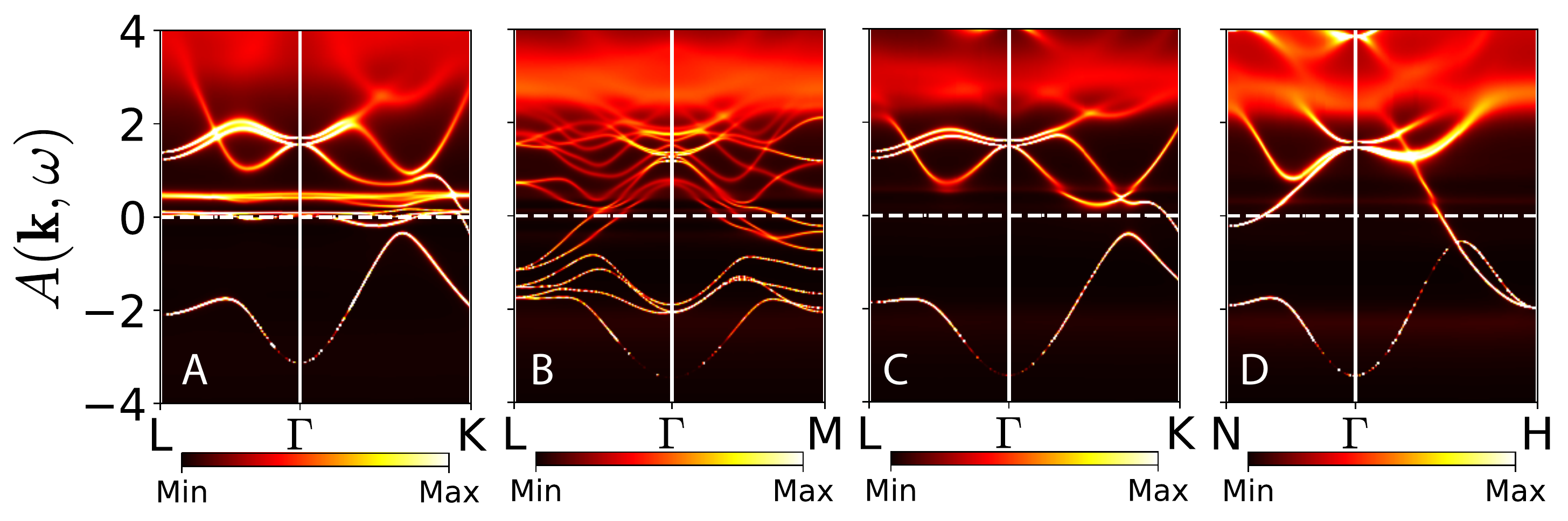}
\caption{(Color online). Momentum-resolved spectral functions $A(\mathbf{k},\omega)$ of Ce by DFT + DMFT calculations. (a) $\alpha$-Ce. (b) $\beta$-Ce. (c) $\gamma$-Ce. (d) $\delta$-Ce. The horizontal dashed lines denote the Fermi level $E_{\text{F}}$. \label{fig:takw}}
\end{figure*}

\begin{figure*}[ht]
\centering
\includegraphics[width=\textwidth]{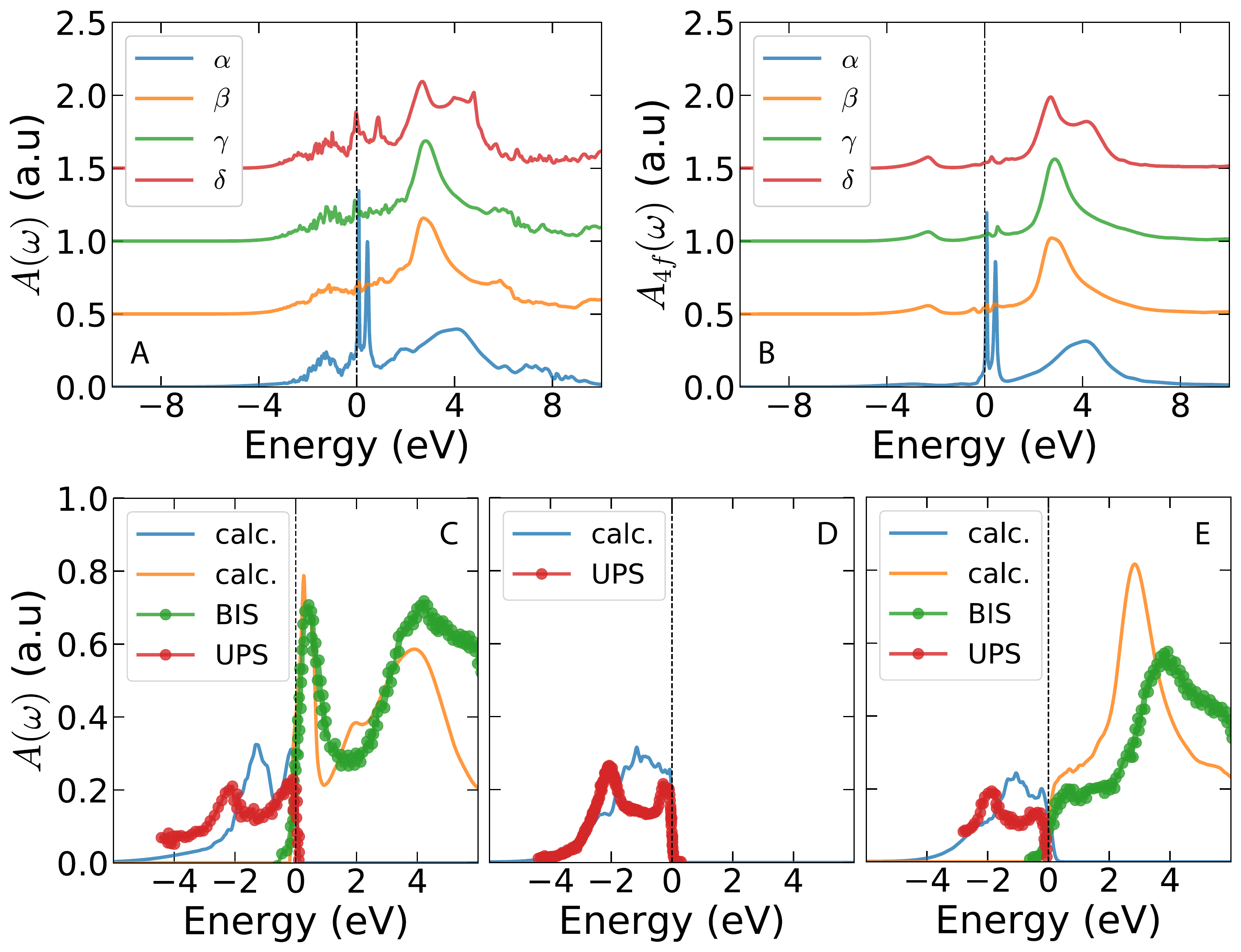}
\caption{(Color online). Density of states of Ce by DFT + DMFT calculations. (a) Total density of states $A(\omega)$. (b) $4f$ partial density of states $A_{4f}(\omega)$. (c)-(e) Comparisons of theoretical and experimental density of states for $\alpha$-, $\beta$-, and $\gamma$-Ce, respectively. In panels (c) and (e), the UPS data (filled red circles) and BIS data (filled green circles) are taken from Ref.~[\onlinecite{PhysRevB.29.3028}] and Ref.~[\onlinecite{PhysRevB.28.7354}], respectively. In panel (d), the experimental data are taken from Ref.~[\onlinecite{PhysRevB.58.3682}]. The Fermi levels $E_{\text{F}}$ are represented by vertical dashed lines. Note that the spectral data have been rescaled and normalized for a better visualization. \label{fig:tdos}}
\end{figure*}

\begin{figure}[ht]
\centering
\includegraphics[width=\columnwidth]{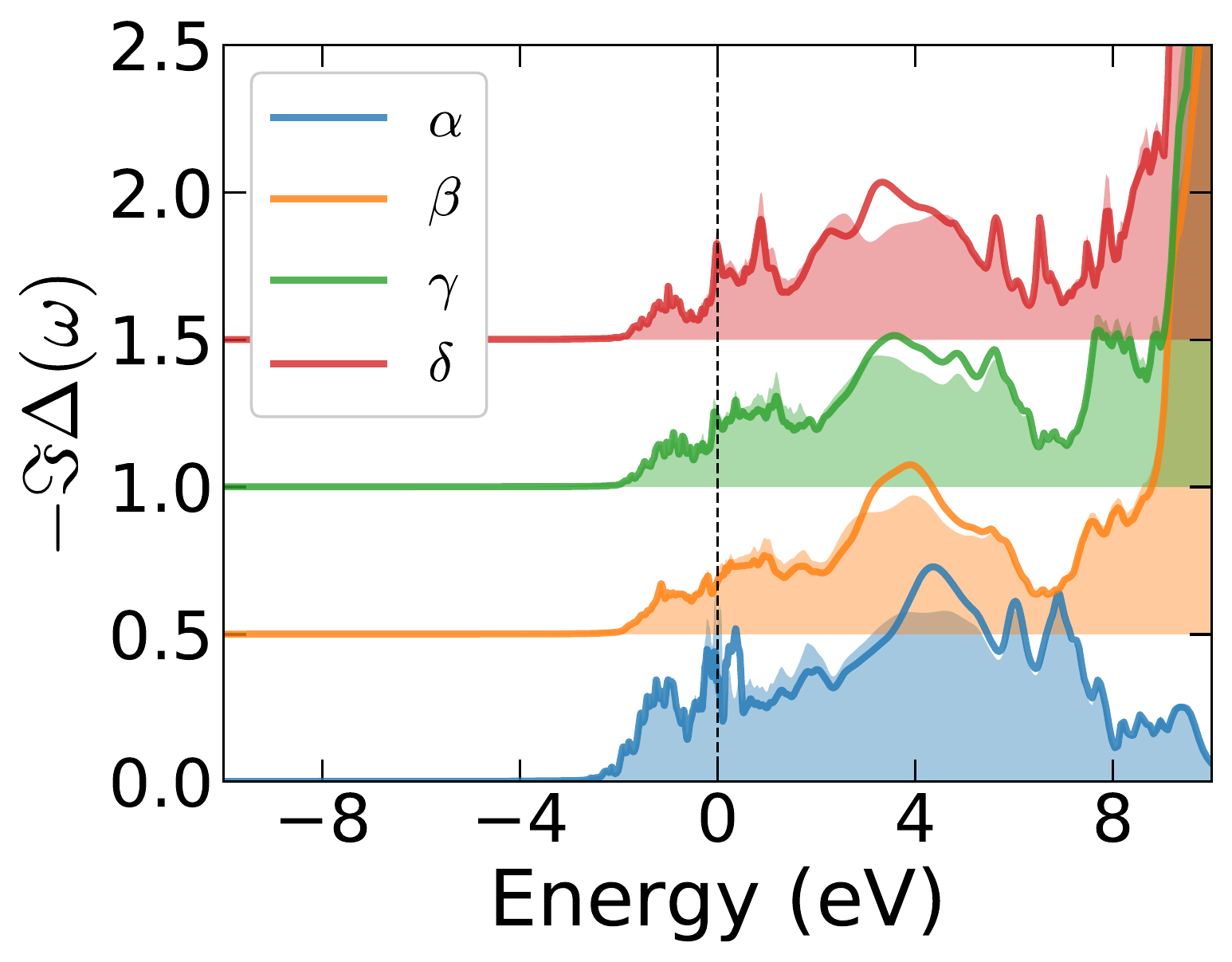}
\caption{(Color online). 4$f$ impurity hybridization functions $-\Im \Delta(\omega)$ of Ce by DFT + DMFT calculations~\cite{hybrid}. The $4f_{5/2}$ and $4f_{7/2}$ components are represented by solid thick lines and color-filled areas, respectively. The Fermi level $E_{\text{F}}$ is denoted by vertical dashed line. Note that the data have been rescaled and shifted upward for a better visualization. \label{fig:tdelta}}
\end{figure}

\begin{figure*}[ht]
\centering
\includegraphics[width=\textwidth]{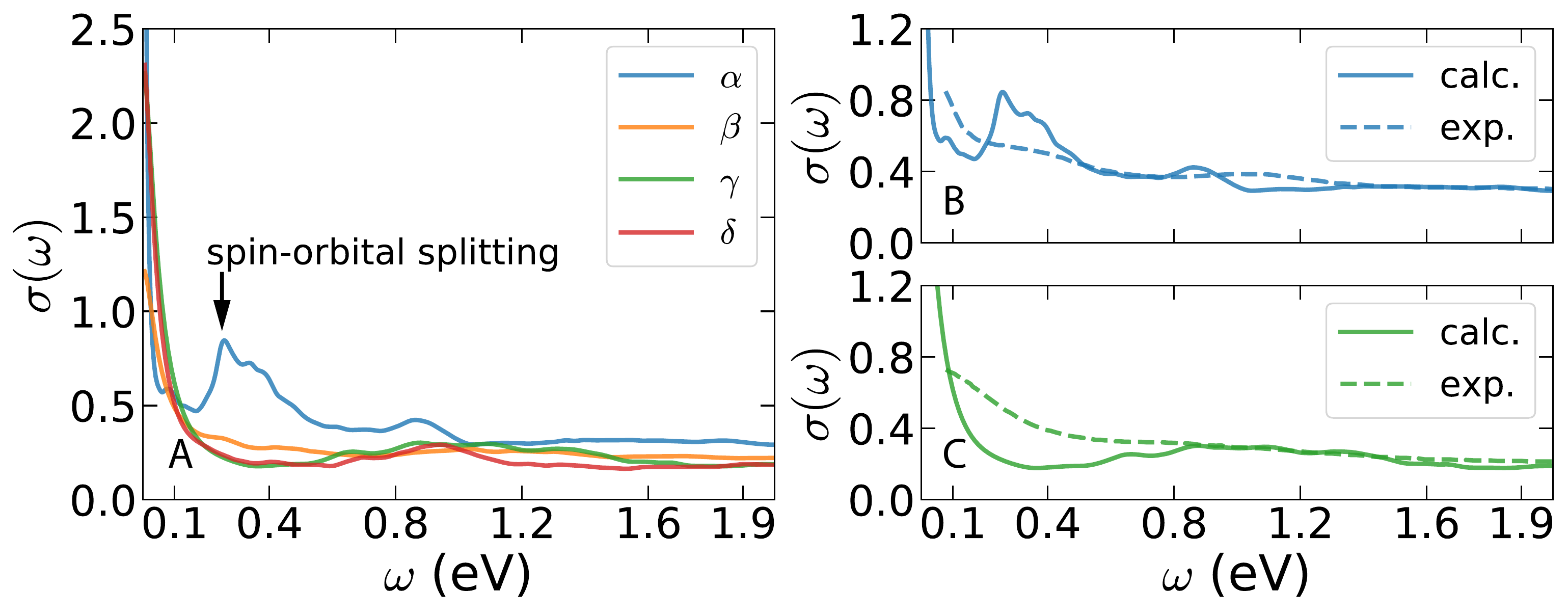}
\caption{(Color online). Optical properties of Ce by DFT + DMFT calculations. (a) $\Re \sigma(\omega)$ in linear coordinate. The unit for $y$-axis is $(\mu \Omega \text{m})^{-1}$. (b)-(c) Calculated and experimental optical conductivities for $\alpha$- and $\gamma$-Ce, respectively. The theoretical data are rescaled by a factor of 0.5 for a better comparison. The experimental data are extracted from Ref.~[\onlinecite{PhysRevLett.86.3407}]. The optical measurements for the $\alpha$ and $\gamma$ phases were done at 5~K and 400~K, respectively.
\label{fig:toptics}}
\end{figure*}

\begin{figure}[ht]
\centering
\includegraphics[width=\columnwidth]{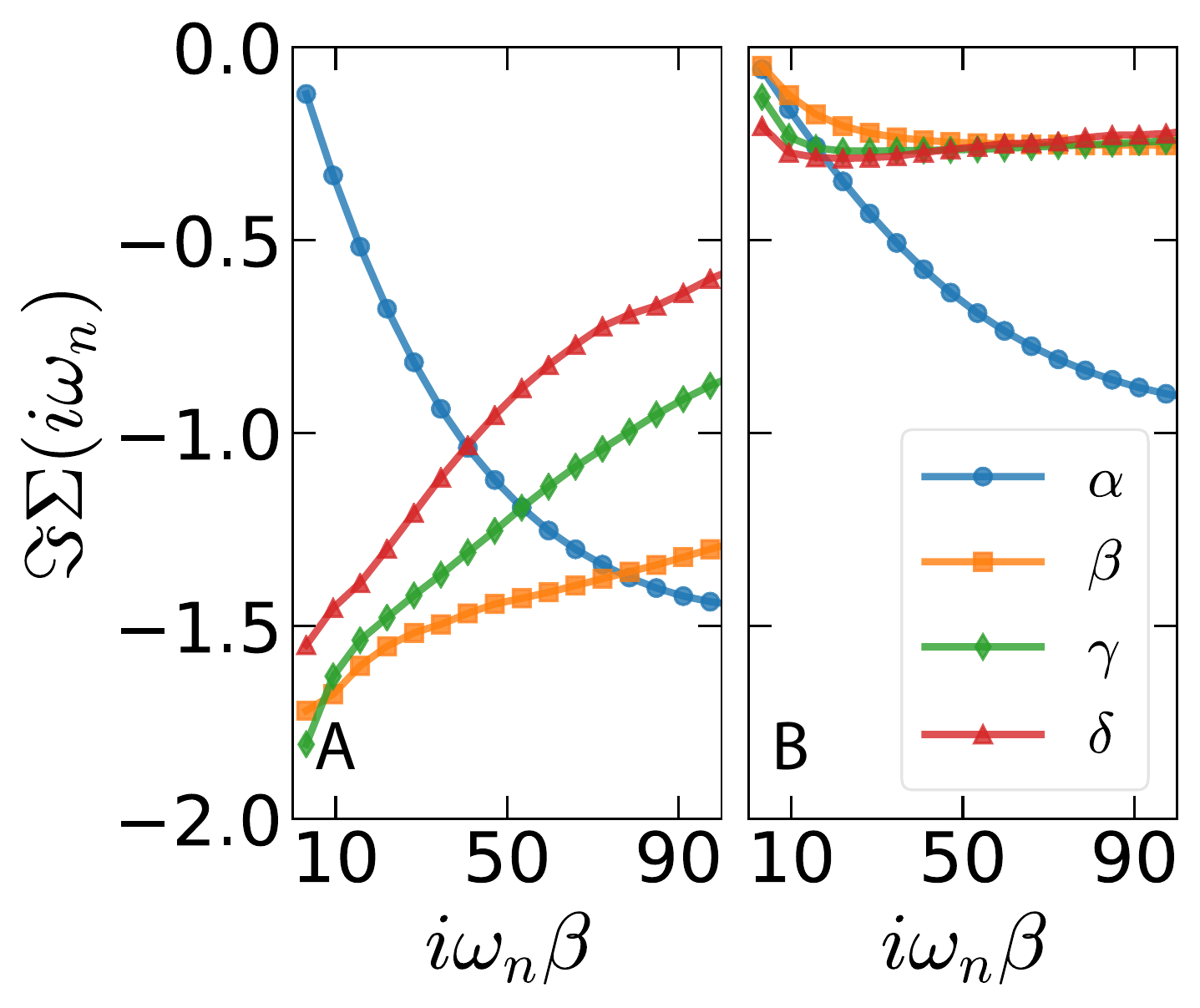}
\caption{(Color online). Matsubara self-energy function $\Sigma(i\omega_n)$ of Ce in low-frequency regime by DFT + DMFT calculations. Only the imaginary parts are shown. (a) $4f_{5/2}$ components. (b) $4f_{7/2}$ components. \label{fig:tsigma}}
\end{figure}

\begin{figure*}[ht]
\centering
\includegraphics[width=\textwidth]{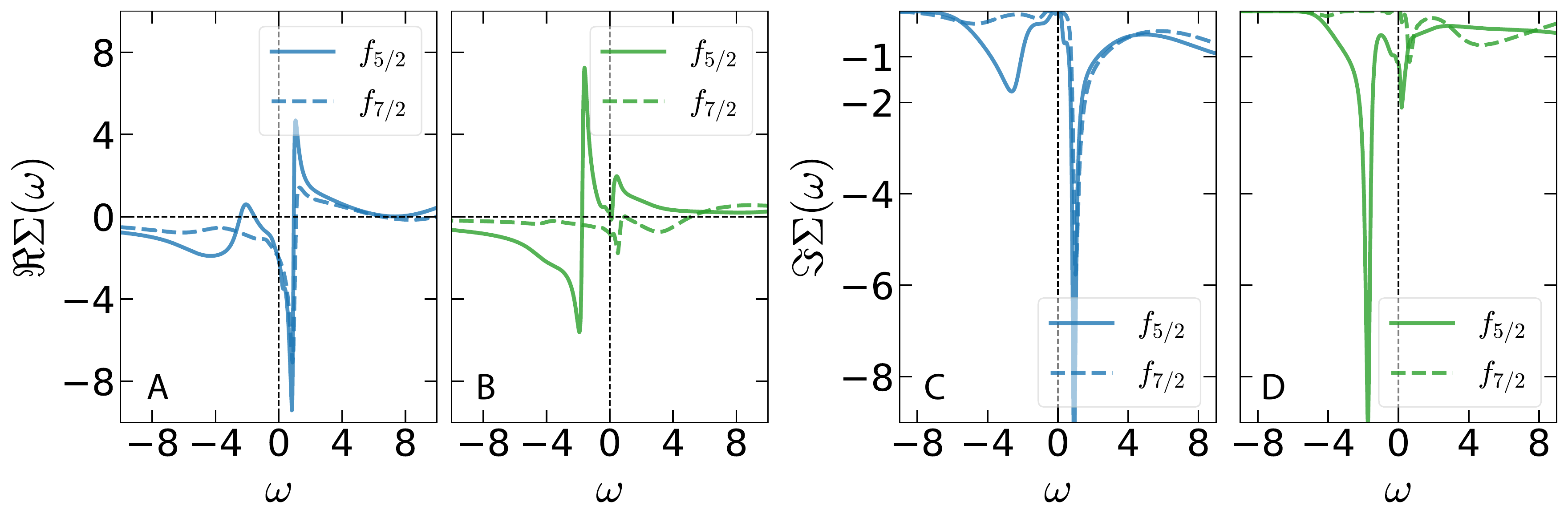}
\caption{(Color online). Real-frequency self-energy functions $\Sigma(\omega)$ (Left panel: real parts. Right panel: imaginary parts) for $\alpha$- and $\gamma$-Ce by DFT + DMFT calculations. (a) and (c) $\alpha$-Ce. (b) and (d) $\gamma$-Ce. \label{fig:tsigma_}}
\end{figure*}

\begin{figure*}[ht]
\centering
\includegraphics[width=\textwidth]{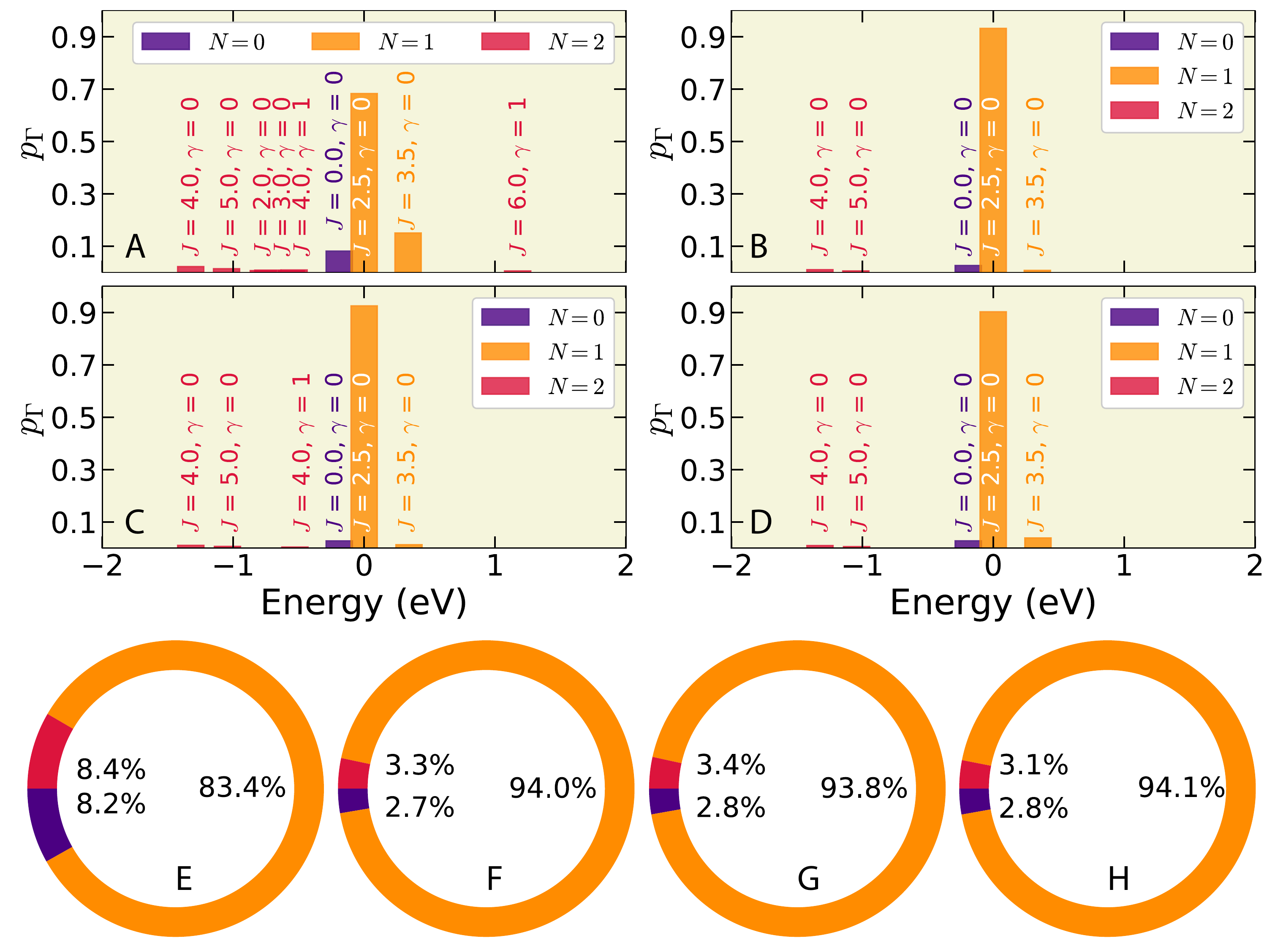}
\caption{(Color online). (a)-(d) Valence state histograms of $\alpha$-, $\beta$-, $\gamma$-, and $\delta$-Ce by DFT + DMFT calculations. Here we used three good quantum numbers to label the atomic eigenstates. They are $N$ (total occupancy), $J$ (total angular momentum), and $\gamma$ ($\gamma$ stands for the rest of the atomic quantum numbers, such as $J_z$). Note that the contribution from $N = 3$ atomic eigenstates is too trivial to be visualized in these panels. (e)-(h) Probabilities of $4f^{0}$ (violet), $4f^{1}$ (orange), and $4f^{2}$ (red) configurations for $\alpha$-, $\beta$-, $\gamma$-, and $\delta$-Ce by DFT + DMFT calculations. \label{fig:tprob}}
\end{figure*}

\subsection{Bulk properties}

Basically, most of the calculated results presented below can be viewed as critical predictions. Are the computational methods and parameters we adopted in the present work reasonable? Are the calculated results reliable? In this and the following subsections, we will compare our results with the experimental data if available to answer these important questions. These comparisons will not only demonstrate the feasibility of the DFT + DMFT approach~\cite{RevModPhys.78.865,RevModPhys.68.13} in the studies of $f$ electronic materials (such as actinides and lanthanides), but also ensure a firm base for future applications of the current predictions.

At first, let's focus on the bulk properties of Ce. The crystal structures of cerium's allotropes have been determined experimentally (see Fig.~\ref{fig:tstruct}). Their crystal volumes, bulk modulus, and (a few) elastic properties have been measured as well~\cite{PhysRevB.19.5746,PhysRevB.25.6485,krisch:2011,huang:2007}. Theoretically, we can use a (semi-)empirical equation of states (EOS) to fit the energy-volume curve of materials to extract the required equilibrium lattice constants $a_0$ and bulk modulus $B$. Here, we take $\gamma$-Ce as an example to illustrate that the DFT + DMFT method can yield a correct description for the bulk properties of $f$ electronic materials. Firstly, we changed the lattice constants of $\gamma$-Ce manually. Then we carried out charge fully self-consistent DFT + DMFT calculations to derive the total energies~\cite{PhysRevLett.115.256402,PhysRevB.81.195107} for given crystal volumes. Finally, we used the Birch-Murnaghan EOS~\cite{PhysRev.71.809} to fit the data. The extracted $a_0$ and $B$, together with the other experimental and theoretical values, are collected in Table~\ref{tab:bulk}. Clearly, our results are rather accurate and agree quite well with the experimental data~\cite{olsen:1985}. The error bars for $a_0$ and $B$ are less than 1.0\% and 1.5\%, respectively. We find that only the DFT + DMFT method can reproduce the equilibrium lattice constants and bulk modulus of $\gamma$-Ce at the same time. The other theoretical methods, such as DFT + $U$~\cite{amadon:2012,Hu2011669,PhysRevB.77.155104}, PBE0~\cite{PhysRevB.93.075153}, and SIC-LSDA~\cite{PhysRevB.53.4275}, usually overestimate the bulk modulus and underestimate the equilibrium lattice constants.

\subsection{Momentum-resolved spectral functions}

Next, we endeavoured to calculate the momentum-resolved spectral functions $A(\mathbf{k},\omega)$ by using Eq.~(\ref{eq:akw}) along the high-symmetry lines in the irreducible Brillouin zone. The calculated results are illustrated in Figure~\ref{fig:takw}. 

The $\alpha$ and $\gamma$ phases share the same crystal structures (Face-centered cubic, see Fig.~\ref{fig:tstruct}). The only difference is that $\gamma$-Ce has larger lattice constants and stabilizes at higher temperature~\cite{alex:2012,koskenmaki1978337}. Therefore it would be very interesting to compare their momentum-resolved spectral functions in depth. Below the Fermi level ($\omega < 0$), both phases exhibit almost identical band structures, which mostly belong to the $spd$ conduction states. Above the Fermi level ($\omega > 0$), the two phases show distinct band structures. For the $\alpha$ phase, there exist intense and almost flat bands in the vicinity of the Fermi level (about 0.02 eV and 0.33 eV) which belong to the spin-orbital splitting 4$f$ bands. In this region, there is strong hybridization between the 4$f$ and $spd$ bands. The spread and blurring features around 4.0\ eV are also associated with the 4$f$ bands. For the $\gamma$ phase, the bands that crossing or getting close to the Fermi level exhibit significant dispersions, which stem from the contributions of itinerant $spd$ electrons. There exist flat 4$f$ bands at $\omega \sim$ 0.3 eV as well, but their intensities are fairly weak. Most of the 4$f$ bands for $\gamma$-Ce locate around 3.0 eV, which have lower energies than those of $\alpha$-Ce. This can be explained by the fact that $\alpha$-Ce has a smaller crystal volume. Furthermore, stronger hybridization between 4$f$ and $spd$ conduction states tends to shift the bands to higher energy. 

As for $\beta$-Ce and $\delta$-Ce, their band structures possess similar features with those of $\gamma$-Ce. In other words, the $spd$ conduction states dominate near the Fermi level, the corresponding bands display large dispersions. There are flat and narrow bands near 3.0 eV which are mainly associated with the 4$f$ character.

\subsection{Density of states}

Now let us turn to the total and 4$f$ partial density of states of Ce. The calculated results are shown in Fig.~\ref{fig:tdos}(a)-(b). As for $\alpha$-Ce, spin-orbital splitting yields two sharp and intensive peaks in the vicinity of Fermi level. They are the so-called Kondo resonance peaks or quasi-particle peaks~\cite{PhysRevLett.87.276404,PhysRevB.72.115125,PhysRevB.67.075108}. The low-lying peak belongs to the $4f_{5/2}$ state, and the high-lying peak attributes to the $4f_{7/2}$ state. Actually, the distance between these two peaks is roughly $\Delta_{\text{SO}}$~\cite{cal_soc}. Without spin-orbital splitting, the two peaks will merge into a single peak, which has been validated by the previous DFT + DMFT calculations for Ce~\cite{PhysRevB.72.115125,PhysRevLett.87.276403,PhysRevLett.87.276404,PhysRevB.67.075108}. Besides the Kondo resonance peaks, the spectra of $\alpha$-Ce include well-separated Hubbard bands which are representative for strongly correlated materials~\cite{RevModPhys.68.13}. The lower and upper Hubbard bands locate at -3.0 eV and 4.0 eV, respectively. Since the nominal 4$f$ occupancy is 1.0, the spectral weight for the lower Hubbard band is quite small. Compared to the spectra of $\alpha$-Ce, those of the $\beta$, $\gamma$, and $\delta$ phases of Ce are somewhat different. Their spectra share some common features. At first, the strong Kondo resonance peaks near the Fermi level are absent, which implies that the hybridization between 4$f$ and $spd$ bands is greatly suppressed in this regime. Second, the lower Hubbard bands become more pronounced. Third, the lower and upper Hubbard bands are shifted towards the Fermi level slightly. Most interestingly, though the crystal structures of $\beta$-Ce and $\gamma$-Ce are completely different, their spectra, regardless of whether total or 4$f$ partial density of states, show amazing similarity. In a short conclusion, the spectral functions of $\alpha$-Ce and $\gamma$-Ce are consistent with the previous DFT + DMFT results~\cite{PhysRevLett.87.276403,PhysRevLett.87.276404,PhysRevB.67.075108,PhysRevB.72.115125,PhysRevB.91.161103}. They are compatible with the Mott transition picture~\cite{bj:1974,PhysRevLett.74.2335} as well as the Kondo volume collapse picture~\cite{PhysRevLett.49.1106,PhysRevB.46.5047} for the Ce $\alpha-\gamma$ phase transition.

Next, we would like to make a detailed comparison between the calculated and the experimental spectra. Let us examine the spectral functions of $\alpha$- and $\gamma$-Ce at first. Since the evolution of spectral functions during the iso-structural Ce $\alpha-\gamma$ transition is very helpful to solve the long-standing controversy about the underlying phase transition mechanism, it has been extensively studied by using the high-resolution photoemission spectroscopy~\cite{PhysRevB.29.3028,PhysRevB.28.7354} and the DFT + DMFT method~\cite{PhysRevLett.87.276404,PhysRevB.72.115125,PhysRevLett.87.276403,PhysRevB.67.075108,PhysRevB.91.161103}. In order to be comparable with the experimental data, the obtained $A(\omega)$ must be multiplied by a Fermi-Dirac distribution function at first, and then broadened using a Gaussian function with carefully chosen smearing parameter $\sigma$. In Fig.~\ref{fig:tdos}(c)-(e), we compare our results with the available UPS (ultraviolet photoemission spectra)~\cite{PhysRevB.29.3028} and BIS (bremsstrahlung isochromat spectra)~\cite{PhysRevB.28.7354} spectra. For $\alpha$-Ce, both the UPS and BIS spectra show apparent Kondo resonance peaks, which are the results of the formation of a single state between the unpaired $f$ electron $4f^1$ and the $spd$ conduction electrons. In principle, the Kondo resonance peak at the Fermi level should be separated into two peaks by spin-orbital splitting. Apparently, the peak at $\omega \sim 0.3$ eV in the BIS spectrum is attributed to the contribution of the $4f_{7/2}$ state, while the peak at $\omega \sim 0.0$ eV in the UPS spectrum is mainly made of the $4f_{5/2}$ state. The calculated spectra successfully reproduce these features. The main structure between 2.0 and 6.0 eV in the BIS spectrum is ascribed to the 4$f^{2}$ final state multiples. It is also verified by the present DFT + DMFT calculation. From Fig.~\ref{fig:tdos}(b), we can conclude that this structure is the so-called upper Hubbard band, which resides around 4.0 eV and describes an excitation of a $4f^2$ state. There are two discrepancies between the theoretical and experimental spectra. First, the peak near -2 eV in the UPS spectrum is the lower Hubbard band. In contrast to experiment, the calculated lower Hubbard band appears to be much broader and its center is approximately located at -3 eV. Second, there exists a peak in $\omega \sim -1.5$ eV in the calculated spectrum, but this peak is missing in the UPS spectrum. It is worth mentioning that the 4$f$ hybridization function for $\alpha$-Ce also exhibits a peak in this place (see Fig.~\ref{fig:tdelta}), so we guess this peak is mainly contributed from the $spd$ conduction electrons. For $\gamma$-Ce, the most conspicuous feature is the absence of sharp Kondo resonance peak, which implies a much lower Kondo temperature than $\alpha$-Ce~\cite{PhysRevLett.87.276404,PhysRevLett.87.276403}. Our calculated spectra correctly reproduce this feature again. As for the lower and upper Hubbard bands, our results agree quite well with the experiment. Similar to $\alpha$-Ce, the peak near -1.5 eV has not been observed in the experimental spectrum. In a word, the calculated spectra are roughly consistent with the experiments~\cite{PhysRevB.29.3028,PhysRevB.28.7354}. The discrepancies could be easily interpreted by the uncertainty in the Coulomb interaction~\cite{Ce_U} or the use of oversimplified double counting term~\cite{jpcm:1997}.

Since $\beta$-Ce stabilizes in a narrow region in the $P-T$ phase diagram of Ce, it has received less attention than $\alpha$- and $\gamma$-Ce~\cite{koskenmaki1978337}. Therefore, very limited experimental results are available for it. And the theoretical studies are also lacking. Because it is the intermediate phase for the temperature-driven Ce $\alpha-\gamma$ transition, it would be valuable to uncover the difference in electronic structures between it and the $\alpha$ (or $\gamma$) phase. The surface states and Fermi surface data for $\beta$-Ce have been recorded using monocrystalline Ce metal film grown on a W(110) surface~\cite{PhysRevB.68.233103}. The experimental data are amazingly in accordance with the momentum-resolved spectral function $A(\mathbf{k},\omega)$ along the $\Gamma-K$ line in the Brillouin zone. As for the density of states, the fine double-peak structures in the Kondo resonance peak and the lower Hubbard band between -3 and -2 eV observed in the UPS experiment~\cite{PhysRevB.58.3682} are successfully replicated by our calculations [see Fig.~\ref{fig:tdos}(d)]. We also observe a strong peak dominated by the $spd$ conducting electrons around -1 eV, but it has not been observed in experiments. This can be parallel to our findings in $\alpha$- and $\gamma$-Ce.

\subsection{Hybridization functions}

Next, we further computed the impurity hybridization functions -$\Im \Delta (\omega)$~\cite{hybrid} of Ce to explore the hybridization effect between the correlated 4$f$ and itinerant $spd$ electrons. The obtained results (both the $4f_{5/2}$ and $4f_{7/2}$ components) are illustrated in Fig.~\ref{fig:tdelta}. The hybridization function of the $\alpha$ phase is notably different from the others. It generally shows stronger peaks and more weights, especially when $\omega \leq 0.0$~eV and 7.0~eV$\geq \omega \geq 4.0$ eV. On the other hand, the hybridization functions for the $\beta$, $\gamma$, and $\delta$ phases of Ce are quite similar, which exhibit large peaks when $\omega > 8.0$~eV. These results manifest that the $spd$ electrons are not pure spectators in the Ce $\alpha-\gamma$ transition since they are bound to the 4$f$ electrons via the hybridization effect. From $\alpha$- to $\gamma$-Ce, the degree of hybridization decreases, and hence the Kondo temperature is reduced~\cite{PhysRevLett.87.276403,PhysRevLett.87.276404}. In the Mott transition scenario, only the 4$f$ bands are involved in the Ce $\alpha-\gamma$ transition. The 4$f$ electrons evolve from the itinerant state in $\alpha$-Ce to the localized state in $\gamma$-Ce~\cite{bj:1974,PhysRevLett.74.2335}. However, in the Kondo volume collapse scenario, the Ce $\alpha-\gamma$ transition is tuned by modifications in the hybridization of the 4$f$ bands with the $spd$ bands~\cite{PhysRevLett.49.1106,PhysRevB.46.5047}. Clearly, our calculated results endorse the Kondo volume collapse scenario for the Ce $\alpha-\gamma$ transition.

\subsection{Optical properties}

Note that K. Haule \emph{et al.} have studied the optical properties of $\alpha$- and $\gamma$-Ce by employing the DFT + DMFT approach~\cite{PhysRevLett.94.036401,PhysRevB.81.195107}. The calculated results agree quite well with the experiments~\cite{PhysRevLett.86.3407}. In the present work, we tried to reproduce their results, and further computed the optical properties of $\beta$- and $\delta$-Ce which have not been reported publicly. The calculated results are shown in Fig.~\ref{fig:toptics}(a). For $\alpha$-Ce, we observe sharp and narrow Drude peak with a width $\sim$ 0.1 eV at half height. There is a peak around 0.3 eV which is due to excitation across the two Kondo resonance peaks. It also shows a deep ``dip" at $\omega \sim$ 0.2 eV and a broad ``bump" near 0.9 eV. On the contrary, the optical spectra for $\beta$-, $\gamma$-, and $\delta$-Ce are almost featureless. They consist of relatively low and fat Drude peaks (the FWHM is about 0.1 $\sim$ 0.2 eV) on a constant background. These results are in consistent with the previous calculations and experiments~\cite{PhysRevLett.94.036401,PhysRevLett.86.3407}. In principle, the $spd$ conducting electrons will dominate the optical response of Ce, since they have very large band dispersions near the Fermi level (see Fig.~\ref{fig:takw}). In the Mott transition scenario for the Ce $\alpha-\gamma$ transition, the role played by the $spd$ electrons is a spectator~\cite{bj:1974,PhysRevLett.74.2335}, hence the optical spectra for the four phases of Ce should not display appreciable changes. On the other hand, in the Kondo volume collapse model, the $spd$ electrons will hybridize with the $4f$ electrons, and play a vital role in the Ce $\alpha-\gamma$ transition~\cite{PhysRevLett.49.1106,PhysRevB.46.5047}. Upon entering the $\alpha$ phase, the hybridization between the $spd$ electrons with $4f$ electrons increases, which will finally lead to the enlarging of hybridization gap and the sharp Drude peak. Obviously, our calculated results tend to support the Kondo volume collapse model for the description of the Ce $\alpha-\gamma$ transition again.

In the literatures, only the infrared and optical spectroscopy for $\alpha$- and $\gamma$-Ce on thin films was reported~\cite{PhysRevLett.86.3407,PhysRevB.51.17390}. Later K. Haule \emph{et al.}~\cite{PhysRevLett.94.036401} have employed the DFT + DMFT method to study the full temperature dependence of the optical spectra of $\alpha$- and $\gamma$-Ce. They used the one-crossing approximation as quantum impurity solver, and their results compared well with the experiments. In the present work, we used the powerful and advanced CT-HYB quantum impurity solver~\cite{RevModPhys.83.349,PhysRevLett.97.076405,PhysRevB.75.155113} to solve the quantum impurity models and DMFT equations, so better results are naturally expected. Here we will compare the calculated optical conductivity $\sigma(\omega)$ with the existing experiments. The comparisons are illustrated in Fig.~\ref{fig:toptics}(b) and (c). First of all, the magnitude of the calculated conductivity in both phases is approximately double of the measurement~\cite{PhysRevLett.86.3407}. K. Haule \emph{et al.}~\cite{PhysRevLett.94.036401} have suggested that neglecting Hubbard interactions among the $spd$ conduction electrons in the DFT + DMFT calculations would be responsible for this discrepancy. On the other hand, in early optical measurements by Rhee \emph{et al.}~\cite{PhysRevB.51.17390}, the magnitude of optical conductivity is double the results of Ref.~[\onlinecite{PhysRevLett.86.3407}]. They also pointed out that the magnitude of measured conductivity is sensitive to the thickness of the films. Second, the major features in the measured conductivity for both phases are qualitatively reproduced by the present DFT + DMFT calculations. For $\alpha$-Ce, it has a larger optical conductivity throughout the calculated energy range. There are a hybridization ``dip" around 0.2 eV and a mid-infrared peak around 0.9 eV. We confirm that the Drude peak in the $\gamma$ phase is much broader than that in the $\alpha$ phase. Furthermore, the calculated conductivity shows subtle convec upward (downward) shape in the $1.5 \sim 2.0$ eV range for the $\alpha$ ($\gamma$) phase which is similar to the early measured ones and theoretical results obtained by FLAPW method~\cite{PhysRevB.51.17390}. Third, in the calculated conductivity for $\gamma$-Ce, we see a prominent ``dip" near 0.4 eV and a broad peak between 0.5 eV and 1.5 eV. Note that in the results obtained by the FLAPW method, similar features were observed and the positions were almost identical~\cite{PhysRevB.51.17390}. In the previous DFT + DMFT calculations~\cite{PhysRevLett.94.036401}, similar peak was observed, but its center was shifted to 1.5 eV. However, the measured conductivity is almost featureless in this region~\cite{PhysRevLett.86.3407}.

\subsection{Self-energy functions}

Next, let us turn to the self-energy functions, which encode all electronic correlations beyond the DFT. The Matsubara $4f$ self-energy functions for Ce are shown in Fig.~\ref{fig:tsigma}. In the low-frequency region, the self-energy functions of $\alpha$-Ce exhibit completely different behaviors as compared to the other phases. For examples, the self-energy functions of $\alpha$-Ce show remarkable metallic like features. Their $y$-intercepts $\Im \Sigma(i\omega_n \to 0)$ approach zero, which imply small low-energy scattering rate. On the contrary, for the $\beta$-, $\gamma$-, and $\delta$- phases, their self-energy functions ($4f_{5/2}$ components) are insulating like, with large low-energy scattering rate. 

Based on the self-energy data, then we can estimate the quasi-particle weights $Z$ and the electron effective masses $m^{\star}$ by using the following equation~\cite{RevModPhys.68.13}:
\begin{equation}
Z^{-1} = \frac{m^{\star}}{m_e} \approx 1 - \frac{\Im\Sigma(i\omega_0)}{\omega_0},
\end{equation}
where $\omega_0 = \pi / \beta$. The calculated results are collected in the Tab.~\ref{tab:mass}. First, the quasi-particle weights $Z$ are small, which reveals that the 4$f$ electrons in cerium are strongly correlated and the band renormalization for the $4f$ orbitals should be considerable. Second, in $\beta$-, $\gamma$-, and $\delta$-Ce, their quasi-particle weights and electron effective masses show very strong orbital dependence. The $4f_{5/2}$ bands usually exhibit smaller $Z$ and larger $m^{\star}$. Third, in the infrared and optical spectroscopy of $\alpha$-Ce, a Fermi-liquid-like frequency-dependent scattering rate was observed~\cite{PhysRevLett.86.3407}. Thus, $\alpha$-Ce was considered to be reminiscent of a Fermi-liquid state with an effective mass of $20\ m_e$, which is contrary to our calculated results. We note that the infrared and optical experiments for $\alpha$-Ce were done on 5~K, while our calculations were carried out using $T = 116$~K (see Table~\ref{tab:param}). The temperature effect is probably the major cause of deviation. Actually, as already shown in K. Haule \emph{et al.}'s work~\cite{PhysRevLett.94.036401}, upon decreasing temperature, the hybridization gap and the Drude peak are growing. At last, the system will approach to a Fermi-liquid state.

\begin{table}[t]
\caption{Calculated quasi-particle weights $Z$ and electron effective masses $m^{\star}$ for Ce. \label{tab:mass}}
\begin{ruledtabular}
\begin{tabular}{ccccc}
      & \multicolumn{2}{c}{$4f_{5/2}$} & \multicolumn{2}{c}{$4f_{7/2}$} \\
cases & $Z$ & $m^{\star}$ & $Z$ & $m^{\star}$ \\
\hline
$\alpha$-Ce & 0.21 & 04.85$m_e$  & 0.36 & 2.81$m_e$ \\
$\beta$-Ce  & 0.04 & 28.36$m_e$ & 0.57 & 1.77$m_e$ \\
$\gamma$-Ce & 0.08 & 12.50$m_e$ & 0.55 & 1.83$m_e$ \\
$\delta$-Ce & 0.13 & 07.61$m_e$  & 0.29 & 3.41$m_e$ \\
\end{tabular}
\end{ruledtabular}
\end{table}

We can convert Matsubara self-energy function $\Sigma(i\omega_n)$ to real-frequency self-energy function $\Sigma(\omega)$ via analytical continuation procedure~\cite{jarrell}. The results for $\alpha$- and $\gamma$-Ce are shown in Fig.~\ref{fig:tsigma_}. For $\alpha$-Ce, we can see that the real parts of self-energy functions $\Re\Sigma(\omega)$ exhibit quasi-linear behavior around the Fermi level, and the imaginary parts of the self-energy functions $\Im \Sigma(\omega)$ indeed exhibit dips at the Fermi level with tiny values of $\Im \Sigma(0)$. As for $\gamma$-Ce, however, it display a zero-frequency value of the imaginary parts of the self-energy as large as 1.0~eV, demonstrating that the coherence temperature of the $4f_{5/2}$ orbitals has not yet been reached. Note that the real-frequency self-energy functions for $\beta$- and $\delta$-Ce share almost the same features with those of $\gamma$-Ce. These facts suggest again that the 4$f$ electrons in $\alpha$-Ce tend to be delocalized, while those in the other phases tend to be localized. Moreover, the self-energy functions and quasi-particle weights data reveal the electronic correlations for $\beta$-, $\gamma$-, and $\delta$-Ce show stronger orbital dependence. Actually, we think the 4$f$ electrons in these phases are in the orbital selective insulating state, wherein the $4f_{5/2}$ electrons are insulating while the $4f_{7/2}$ electrons remain metallic. We believe this state is common in the other cerium-based heavy fermion materials and intermediate valence compounds. It is an interesting analogy to the well-known orbital selective Mott phase found in the $3d$ transition metal oxides~\cite{Anisimov2002}.

\subsection{Valence state histograms}

In many Ce-based heavy fermion compounds, cerium's 4$f$ electrons usually exhibit fascinating mixed-valence (or equivalently valence fluctuation) behaviors, resulting in non-integer $4f$ occupancy~\cite{PhysRevLett.96.237403,PhysRevB.67.075108}. As for the valence fluctuation in pure Ce metal, unfortunately, it has been rarely studied and reported in the public literatures~\cite{PhysRevB.75.155113}. Currently, the best approach to analyze the valence fluctuation quantitatively is via the valence state histogram (or atomic eigenstate probability), which represents the probability $p_{\Gamma}$ to find a valence electron in a given atomic eigenstate $| \psi_{\Gamma} \rangle$~\cite{shim:2007,PhysRevB.81.195107}. Thanks to the CT-HYB quantum impurity solver, we can extract the valence state histogram~\cite{PhysRevB.75.155113} from its output directly. So, in the present work, we studied the valence fluctuation behavior in Ce rigorously. 

As mentioned before, the atomic eigenstates for $4f$ electrons can be labelled using some good quantum numbers, such as $N$ and $J$. In Fig.~\ref{fig:tprob}, we visualize their histograms for $\alpha$-, $\beta$-, $\gamma$-, and $\delta$-Ce in the (a)-(d) panels, respectively. We find that the $|N = 1, J = 2.5, \gamma = 0 \rangle$ atomic eigenstate is overwhelmingly dominant in these phases. For example, its probability accounts for 68.31\% in the $\alpha$ phase, and even larger than 90\% in $\beta$-, $\gamma$-, and $\delta$-Ce. In the $\alpha$ phase, the probabilities for the $|N = 0, J = 0.0, \gamma = 0\rangle$ and $| N = 1, J = 3.5, \gamma = 0 \rangle$ atomic eigenstates are also considerable, which are 8.16\% and 15.02\%, respectively. However, in the other phases, the probabilities for the two states are trivial. This observation suggests that the 4$f$ electrons in the $\alpha$ phase favor to wander among different atomic eigenstates, instead of being trapped in a given atomic eigenstate.   

From the valence state histogram, the distributions of 4$f$ electronic configurations can be easily computed via the following equation:
\begin{equation}
w(4f^{i}) = \sum_{N}\sum_{J}\sum_{\gamma} \delta(N - i) p_{\Gamma}.
\end{equation}
Here, $w(4f^{i})$ denotes the weight for the $4f^{i}$ (where $i \in [0,3]$) electronic configuration. Next we turn to discuss the distributions of 4$f$ configurations [see Fig.~\ref{fig:tprob}(e)-(h)]. First, we choose the following equation to evaluate the averaged 4$f$ occupancies $n_{4f}$ approximately:
\begin{equation}
 n_{4f} \approx  w(4f^{1}) + 2w(4f^{2}) + 3 w(4f^{3}).
\end{equation}
As is expected, the calculated values are very close to 1.0 for all phases. Second, the ruling configuration is always $4f^{1}$, while the $4f^{3}$ configuration is ignorable~\cite{PhysRevB.67.075108,PhysRevLett.87.276403}. Let us focus on $w(4f^{1})$. We attempted to plot the $w(4f^{1})$ against the unit cell volume $V$ of Ce. Quite surprisingly, we find that $w(4f^{1}) - V$ exhibits a quasi-linear relation. The $w(4f^{1})$ increases monotonically with respect to $V$. We think that the $w(4f^{1})$ can be considered as a quantitative tool to measure the status of 4$f$ electrons of Ce. Clearly, the 4$f$ electrons in the $\alpha$ phase are the most itinerant, with the smallest $w(4f^{1})$ and $V$. On the contrary, in the high-temperature $\delta$ phase, its 4$f$ electrons are the most localized, with the largest $w(4f^{1})$ and $V$. The abnormal volume decrease during the $\beta-\gamma$ phase transition~\cite{koskenmaki1978337} can be also understood by the fact that $\beta$-Ce has a larger $w(4f^1)$ than $\gamma$-Ce. Third, we find that from $\alpha$- to $\beta$-Ce, dramatic change occurs in the relative ratio between different 4$f$ electronic configurations. In $\alpha$-Ce, the proportions for $4f^{0}$, $4f^{1}$, and $4f^{2}$ are 8.2\%, 83.4\%, and 8.4\%, respectively. But in $\beta$-Ce, the proportion for $4f^{1}$ configuration increases up to 94.0\%, those for $4f^{0}$ and $4f^{2}$ configurations decrease to 2.7\% and 3.3\%, respectively. The situations for $\gamma$- and $\delta$-Ce are very similar to $\beta$-Ce. They share almost the same electronic configurations. In other words, $\alpha$-Ce has more 4$f^0$ and 4$f^2$, and less 4$f^1$. These results indicate that owing to strong hybridization effect, the valence fluctuation in $\alpha$-Ce is more prominent than the ones in the other phases. Accordingly, the distribution of 4$f$ configurations in $\alpha$-Ce is more diverse.

The valence state fluctuations in Ce-based heavy fermion materials~\cite{yuan:2016} are always hot topics in the research field of strongly correlated systems, and have attracted many attentions and efforts. However, the valence state fluctuations and 4$f$ electronic configurations in $\beta$- and $\delta$-Ce remain unknown. Our calculated results not only supplement the missing data, but also provide a unified picture for the valence state fluctuations in Ce. Here, we would like to compare the calculated 4$f$ electronic configurations for $\alpha$- and $\gamma$-Ce with the measured values to validate the correctness of our DFT + DMFT calculations. The comparison is shown in Tab.~\ref{tab:prob}. As a useful reference, the previously calculated values obtained with the DFT + DMFT method are also collected and displayed in this table. Obviously, our results are consistent with the experimental values~\cite{PhysRevLett.96.237403} and previous DFT + DMFT results~\cite{PhysRevB.75.155113}, especially for the $\gamma$ phase. The experiments were done by Rueff \emph{et al.} using the resonant inelastic X-ray scattering technology~\cite{PhysRevLett.96.237403}. They carried out measurements at 0, 10, and 20~kbar, the corresponding crystal volumes are 34.4, 27.6, and 26.7 \AA$^3$, respectively. Clearly, at 0~kbar, it is the $\gamma$ phase. At room temperature, the Ce $\gamma-\alpha$ transition occurs at $P_c \sim 8$~kbar~\cite{alex:2012,koskenmaki1978337}. The crystal volume for $\alpha$-Ce is 28.1 \AA$^3$. So the data measured at $P =$ 10~kbar are supposed to be more reasonable for the $\alpha$ phase. They are consistent with our results. The data measured at 20~kbar show big difference which indicates that the distribution of the 4$f$ electronic configurations is very sensitive to the crystal volume.

\begin{table}[t]
\caption{Calculated and experimental weights of the 4$f$ electronic configurations for $\alpha$- and $\gamma$-Ce. \label{tab:prob}}
\begin{ruledtabular}
\begin{tabular}{ccccccc}
       & \multicolumn{3}{c}{$\alpha$-Ce} & \multicolumn{3}{c}{$\gamma$-Ce} \\
method & $4f^{0}$ & $4f^{1}$ & $4f^{2}$ & $4f^{0}$ & $4f^{1}$ & $4f^{2}$ \\
\hline
DFT + DMFT\footnotemark[1] & 8.2\%  & 83.4\%  & 8.4\%  & 2.8\%  & 93.8\%  & 3.4\% \\
DFT + DMFT\footnotemark[2] & 11.8\% & 77.1\%  & 11.1\% & 1.3\%  & 93.9\%  & 4.8\% \\
DFT + DMFT\footnotemark[3] & 28.8\% & 66.4\%  & 4.8\%  & 2.3\%  & 94.8\%  & 2.9\% \\
DFT + DMFT\footnotemark[4] & 12.6\% & 82.9\%  & 4.4\%  & 1.5\%  & 94.3\%  & 4.2\% \\
LSDA\footnotemark[5] & 15.6\%  & 80.8\%  & 2.6\%  & 4.3\%  & 94.4\%  & 1.3\% \\
RIXS\footnotemark[6] &         &         &        & 5.1\%  & 92.5\%  & 2.4\%  \\
RIXS\footnotemark[7] & 7.7\%   & 89.6\%  & 2.7\%  &        &         &        \\
RIXS\footnotemark[8] & 21.9\%  & 74.7\%  & 3.4\%  &        &         &        \\
\end{tabular}
\end{ruledtabular}
\footnotetext[1]{The present work.}
\footnotetext[2]{With Hirsch-Fye quantum Monte Carlo impurity solver. See Ref.~[\onlinecite{PhysRevB.67.075108}].}
\footnotetext[3]{With CT-HYB quantum Monte Carlo impurity solver. See Ref.~[\onlinecite{PhysRevB.75.155113}].}
\footnotetext[4]{With non-crossing approximation impurity solver. See Ref.~[\onlinecite{PhysRevLett.87.276403}].}
\footnotetext[5]{See Ref.~[\onlinecite{PhysRevLett.87.276403}].}
\footnotetext[6]{Resonant inelastic X-Ray scattering. $P = 0$~kbar. See Ref.~[\onlinecite{PhysRevLett.96.237403}].}
\footnotetext[7]{Resonant inelastic X-Ray scattering. $P = 10$~kbar. See Ref.~[\onlinecite{PhysRevLett.96.237403}].}
\footnotetext[8]{Resonant inelastic X-Ray scattering. $P = 20$~kbar. See Ref.~[\onlinecite{PhysRevLett.96.237403}].}
\end{table}


\section{Discussion\label{sec:discuss}}

\begin{figure}[t]
\centering
\includegraphics[width=\columnwidth]{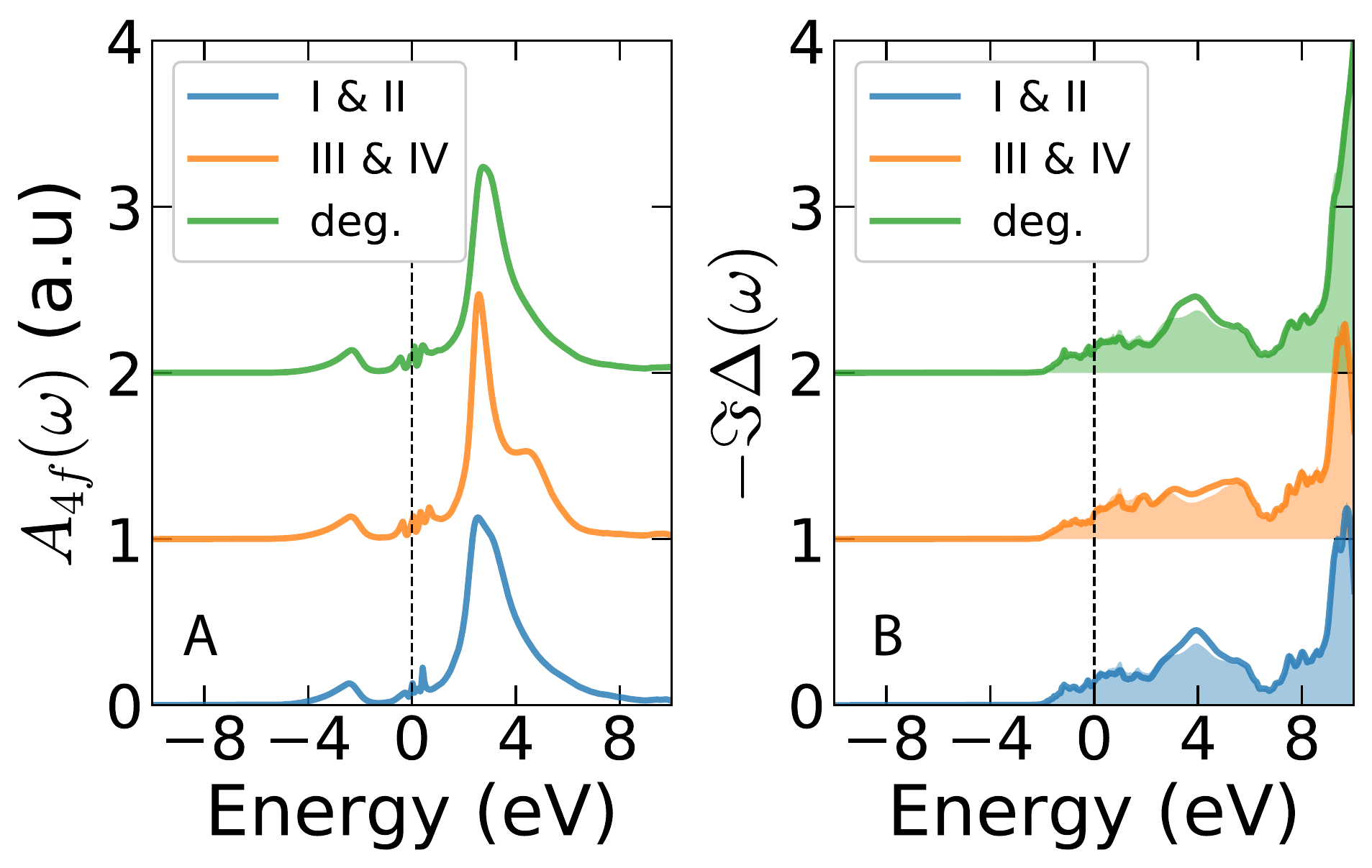}
\caption{(Color online). Site-resolved electronic structures in $\beta$-Ce by DFT + DMFT calculations. (a) $4f$ partial density of states $A_{4f}(\omega)$. (b) Imaginary parts of hybridization functions $-\Im \Delta(\omega)$. The $4f_{5/2}$ and $4f_{7/2}$ components are represented by solid thick lines and color-filled areas, respectively. \label{fig:tsite}}
\end{figure}

The crystal structure of $\beta$-Ce is double hexagonal close-packed. It has four Ce atoms in the unit cell, we label them as Ce$_{\text{I}}$, Ce$_{\text{II}}$, Ce$_{\text{III}}$, and Ce$_{\text{IV}}$ to facilitate further discussion (see Fig.~\ref{fig:tstruct}). Since Ce$_{\text{I}}$ and Ce$_{\text{II}}$, Ce$_{\text{III}}$ and Ce$_{\text{IV}}$ are equivalent atoms, respectively, in principle there are two non-equivalent Ce atoms with fractional coordinates (0.0,0.0,0.0) and (1/3,2/3,1/4) in $\beta$-Ce~\cite{koskenmaki1978337}. In the framework of DFT + DMFT method, each non-equivalent atom is described by a unique quantum impurity model, which should be solved individually in a self-consistent manner~\cite{RevModPhys.78.865,RevModPhys.68.13}. As was mentioned before, to solve the quantum impurity problems for $f$-electron systems using the CT-HYB quantum impurity solver (which is already the most powerful impurity solver so far) is extremely memory-consuming and time-consuming owing to the exponentially increasing Hilbert space and severe negative sign problem~\cite{zhu:2013,RevModPhys.83.349}. For this reason, in the present work, we have to restrict ourselves to consider only the completely degenerated Ce atoms. In other words, we assume that the Ce atoms are all equivalent, irrespective of the original crystal structure and symmetry. This assumption simplifies the calculations greatly, but undoubtedly leads to deviations to some extent. Here, we would like to discuss the site dependence of $4f$ electronic structures in $\beta$-Ce, and clarify the consequence of applying this assumption.

We carried out two benchmark calculations to examine the fine electronic structures of each non-equivalent Ce atoms in $\beta$-Ce. First, the four Ce atoms were forced to be degenerated, the calculated results have been shown in previous sections [see Fig.~\ref{fig:tdos} and Fig.~\ref{fig:tdelta}]. Second, we considered two non-equivalent Ce atoms, and then performed calculations using the same computational parameters. Since the bond distance for Ce$_{\text{I}}$-Ce$_{\text{II}}$ (5.9285~\AA) is somewhat smaller than the one of Ce$_{\text{III}}$-Ce$_{\text{IV}}$ (6.2979~\AA), we speculate that $\beta$-Ce has site-dependent 4$f$ electronic structures. Specially speaking, $\beta$-Ce probably shows a site-selective 4$f$ localized state, in which the 4$f$ electrons in the Ce$_{\text{I}}$ and Ce$_{\text{II}}$ sites are itinerant-like, while in Ce$_{\text{III}}$ and Ce$_{\text{IV}}$ sites are localized-like. However, the calculated results are contrary to what we expected. In Fig.~\ref{fig:tsite}(a), we find that the $A_{4f}(\omega)$ for the Ce atoms at the I-IV sites are very similar. The difference is rather trivial. The degenerated 4$f$ partial density of states can be seen as an envelope line or an average of the non-degenerated ones. In addition to $A_{4f}(\omega)$, we also compare the site-resolved and degenerated hybridization functions in Fig.~\ref{fig:tsite}(b). They also exhibit similar features. All these results suggest that the 4$f$ electronic structures in $\beta$-Ce are weakly site-dependent. It means that our assumption is quite reasonable, and won't change the final results too much. Zhu \emph{et al.}~\cite{zhu:2013} have revealed weak site dependence in the electronic structures of $\alpha$-Pu. We thus expect that in the low-symmetry phases of 4$f$ and $5f$ systems, the electronic structures would exhibit some kind of non-trivial site-dependent features. This is an open and interesting question. In the future calculations for the other low-symmetry 4$f$- or 5$f$ strongly correlated metals, we will take it into serious consideration if computational conditions permit. 


\section{Concluding remarks\label{sec:summary}}

In the present paper, we employed the \emph{ab initio} many-body approach, namely the charge fully self-consistent DFT + DMFT method, to study the electronic structures and optical properties of 4$f$ strongly correlated metal Ce. We endeavored to calculate the momentum-resolved spectral functions, total and $4f$ partial density of states, valence state histograms, and optical conductivities for cerium's allotropes under ambient pressure. The calculated results agree quite well with the available experimental results. Most of the results presented in this paper can be viewed as critical predictions and require further experimental or theoretical examinations. Besides, the other major findings of this work are as follows: (1) We confirmed again that the 4$f$ electrons in $\alpha$-, $\beta$-, $\gamma$-, and $\delta$-Ce are strongly correlated. The band and electron mass renormalizations are not only remarkable, but also strongly orbital-dependent. (2) The calculated spectral functions and hybridization functions prefer the Kondo volume collapse scenario for the 4$f$ electronic structure transition between the $\alpha$ and $\gamma$ phases~\cite{PhysRevLett.49.1106,PhysRevB.46.5047}. (3) The valence state fluctuation in $\alpha$-Ce is much stronger than the others, and its optical conductivity shows strong Drude peak and non-Fermi-liquid behavior in the low-frequency region. (4) The 4$f$ electronic structures for $\beta$-, $\gamma$-, and $\delta$-Ce are very similar.  In $\beta$-Ce, the site dependence of electronic structure is very weak. The fascinating site selective localized state is not observed.

Our work suggest that the state-of-the-art DFT + DMFT method can be applied to study the complex 4$f$ strongly correlated electron systems quantitatively. It can not only reproduce the experimental results, but also discover some new physics. We believe that our calculated results are useful supplements to the experiments, and will shed new light on the \emph{ab initio} calculations for lanthanides and actinides. Note that besides Ce, some light and middle 5$f$ actinide metals (such as Pa, U, Np, and Pu) show complex phase diagrams and phase transitions as a function of temperature. On the other hand, many heavy actinide metals (such as Am, Cm, and Bk) also show various crystallographic phases under pressure~\cite{RevModPhys.81.235}. The electronic structures for most of them remain unclear. It would be highly promising to apply the powerful DFT + DMFT approach to explore their intriguing properties in the near future.


\begin{acknowledgments}
We acknowledge the fruitful discussions with Prof. Kristjan Haule and Dr. Xie-Gang Zhu. This work was supported by the Natural Science Foundation of China (Grant No.~11504340 and No.~11704347), the Foundation of President of China Academy of Engineering Physics (Grant No.~YZ2015012). The DFT + DMFT calculations were performed on the Delta cluster (in the Institute of Physics, CAS, China).
\end{acknowledgments}


\bibliography{ce}

\begin{thebibliography}{73}
\bibitem [{cal()}]{cal_soc}%
  \BibitemOpen
  \href@noop {} {}\bibinfo {note} {The splitting between the $4f_{5/2}$ and
  $4f_{7/2}$ impurity levels is $\Delta_{\text{SO}} = 7/2 \lambda_{\text{SO}}$.
  Since $\lambda_{\text{SO}} \approx 0.0953$ in Ce (see Table~\ref{tab:param}),
  $\Delta_{\text{SO}} \approx 0.33$ eV.}\BibitemShut {Stop}%
\bibitem [{Ce_()}]{Ce_U}%
  \BibitemOpen
  \href@noop {} {}\bibinfo {note} {Note that the dynamical screening effect in
  Ce is remarkable, a static approximation for the screened interaction $U$ in
  Ce (especially in $\alpha$-Ce) is too
  drastic~\cite{PhysRevB.89.125110}.}\BibitemShut {Stop}%
\end{thebibliography}

\end{document}